\documentclass[useAMS,usenatbib]{mn2e}

 \usepackage{graphicx}
 \usepackage{comment}
 \usepackage{color}
 \usepackage{epstopdf}
 
\title[IMF--SFR Relationship]{Galaxy And Mass Assembly (GAMA): The star formation rate dependence of the stellar initial mass function}
\author[Gunawardhana, Hopkins, Sharp et al.]{M. L. P. Gunawardhana$^{1,2,3}$\thanks{E-mail:
mlpg@physics.usyd.edu.au}, A. M. Hopkins$^{2}$\thanks{E-mail: ahopkins@aao.gov.au}, R. G. Sharp$^{2}$, S. Brough$^2$, E. Taylor$^3$, \newauthor J. Bland--Hawthorn$^{3}$, C. Maraston$^{4}$, R. J. Tuffs$^{5}$, C. C. Popescu$^{6}$, D. Wijesinghe$^3$, \newauthor  D. H. Jones$^2$, S. Croom$^{3}$, E. Sadler$^{3}$, S. Wilkins$^{7}$, S. P. Driver$^8$, J. Liske$^9$, P. Norberg$^{10}$, \newauthor I. K. Baldry$^{11}$, S. P. Bamford$^{12}$, J. Loveday$^{13}$, J. A. Peacock$^{10}$, A. S. G. Robotham$^{8}$, \newauthor D. B. Zucker$^{1,2}$, Q. A. Parker$^{1,2}$, C. J. Conselice$^{12}$, E. Cameron$^{8,14}$, C. S. Frenk$^{15}$,  \newauthor D. T. Hill$^8$, L. S. Kelvin$^8$, K. Kuijken$^{16}$, B. F. Madore$^{17}$, B. Nichol$^{5}$, H. R. Parkinson$^{10}$, \newauthor K. A. Pimbblet$^{18}$, M. Prescott$^{11}$,  W. J. Sutherland$^{19}$,  D. Thomas$^{5}$, E. van Kampen$^9$ \\
$^1$Department of Physics and Astronomy, Macquarie University, Sydney, NSW 2109, Australia\\
$^2$Australian Astronomical Observatory, Epping, NSW 1710, Australia\\
$^3$Sydney Institute for Astronomy, School of Physics, University of Sydney, NSW 2006, Australia\\
$^{4}$Institute of Cosmology and Gravitation (ICG), Dennis Sciama Building, Burnaby Road, University of Portsmouth, PO13FX\\
$^{5}$Max Planck Institute for Nuclear Physics (MPIK), Saupfercheckweg 1, D-69117 Heidelberg, Germany\\
$^{6}$Jeremiah Horrocks Institute, University of Central Lancashire, Preston PR1 2HE\\
$^{7}$University of Oxford, Department of Physics, Denys Wilkinson Building, Keble Road, OX1 3RH, U.K.\\
$^8$Scottish UniversitiesÕ Physics Alliance (SUPA), School of Physics and Astronomy, University of St. Andrews, North Haugh, St. Andrews,\\\ \ Fife, KY16 9SS\\
$^9$European Southern Observatory, Karl-Schwarzschild-Str. 2, 85748 Garching, Germany\\
$^{10}$SUPA, Institute for Astronomy, University of Edinburgh, Royal Observatory, Blackford Hill, Edinburgh EH9 3HJ\\ $^{11}$Astrophysics Research Institute, Liverpool John Moores University, Twelve Quays House, Egerton Wharf, Birkenhead CH41 1LD\\
$^{12}$Centre for Astronomy and Particle Theory, University of Nottingham, University Park, Nottingham NG72RD\\ $^{13}$Astronomy Centre, University of Sussex, Falmer, Brighton, BN1 9QH\\
$^{14}$Department of Physics, Swiss Federal Institute of Technology (ETH-Z$\ddot{u}$rich), CH-8093Z$\ddot{u}$rich, Switzerland\\
$^{15}$Institute for Computational Cosmology, Department of Physics, Durham University, South Road, Durham DH1 3LE\\ $^{16}$P.O. Box 9513, NL-2300 RA  Leiden, The Netherlands\\
$^{17}$Observatories of the Carnegie Institution for Science, 813 Santa Barbara Street, CA 91101 USA\\
$^{18}$School of Physics, Monash University, Clayton, Victoria 3800, Australia\\
$^{19}$Astronomy Unit, Queen Mary University London, Mile End Rd, London E14NS}

\begin{document}

\date{Accepted 2011 March 28}

\pagerange{\pageref{firstpage}--\pageref{lastpage}} \pubyear{2002}

\maketitle

\label{firstpage}

\begin{abstract}
The stellar initial mass function (IMF) describes the distribution in stellar masses produced from a burst of star formation. For more than fifty years, the implicit assumption underpinning most areas of research involving the IMF has been that it is universal, regardless of time and environment. We measure the high--mass IMF slope for a sample of low--to--moderate redshift galaxies from the Galaxy And Mass Assembly survey. The large range in luminosities and galaxy masses of the sample permits the exploration of underlying IMF dependencies. A strong IMF--star formation rate dependency is discovered, which shows that highly star forming galaxies form proportionally more massive stars (they have IMFs with flatter power--law slopes) than galaxies with low star formation rates. This has a significant impact on a wide variety of galaxy evolution studies, all of which rely on assumptions about the slope of the IMF. Our result is supported by, and provides an explanation for, the results of numerous recent explorations suggesting a variation of or evolution in the IMF.
\end{abstract}
\begin{keywords}
galaxies--stellar initial mass function: galaxies--formation and evolution
\end{keywords}

\section{Introduction}

The stellar initial mass function (IMF) is an empirical power--law relation describing the distribution of stellar masses formed in a single episode of star formation. This initial stellar mass distribution has often been assumed to be universal. This is perhaps the most fundamental assumption used in all galaxy formation and evolution studies. The IMF is the bridge between the massive stars, measurable through tracers such as  H$\alpha$, ultraviolet, far--infrared and radio luminosity, and the low mass stars, which form the bulk of the stellar mass in galaxies \citep{Kennicutt98}.  
The IMF is intimately involved in many aspects of the modeling of galaxy evolution. Some examples include models of turbulent fragmentation and collapse of gas clouds that form sub--stellar to super--stellar objects \citep{Nakamura01}, the numerical study of supersonic hydrodynamics and magnetohydrodynamics of turbulence \citep{Padoan07}, gradual processes behind building of a galaxy \citep{Gibson97}, the reionisation of the intergalactic medium at high redshift ($z>6$) \citep{Chary08}, the relationship between stellar mass and star formation rate (SFR) \citep{Dave08}, evolution in colour and mass--to--light ratio of galaxies \citep{V08}, models of heavy element production, chemical enrichment and evolution in galaxies \citep{Calura09} from the death of stars, the fraction of stars that form black holes \citep{Fryer03} and many other evolutionary processes.

The IMF is often parameterised as one or more power laws, describing the number of stars within a given mass interval, $\frac{dN}{dM}\,\propto\,m^{\alpha}$ with $\alpha$ defining the slope for the mass range of interest \citep{Salpeter55, BG03}. Other widely adopted functional forms for the IMF include a lognormal form for the low mass regime with a power--law tail for high masses \citep{C01}. From star counts of local resolved stellar populations, the IMF is measured to have a high-mass ($m>0.5 \,M_{\odot}$) slope of $\alpha \approx -2.35$  \citep{Salpeter55, S86}, also called the Salpeter slope, although variations of this slope are also reported \citep{K93, Kroupa01, MS79}. An ``integrated galaxy" IMF (IGIMF) has recently been proposed for the interpretation of the galaxy--wide IMF properties. Whether the IGIMF is universal \citep{E06} or differs from the star cluster IMF \citep{WK05a} is again a much debated subject. The dependence of cluster formation on the SFR of a galaxy is argued to give rise to a varying IGIMF \citep{WK05a}, although the underlying IMF may still be universal. In almost all cases, the high--mass IMF slope is modelled with a power--law type behaviour and a multi-part IMF expression best describes the stellar luminosity function in the solar neighbourhood \citep{S86, K93}. Therefore, a multi--part power law is used to obtain the results presented here. 
For the purpose of this investigation, we primarily use a 2--part power law with a Salpeter high--mass ($>0.5\,M_{\odot}$) slope and $\alpha=-1.3$ low--mass ($0.1\leq M/M_{\odot} \leq0.5$) slope. The use of other popular functional forms of the IMF \citep{MS79, S86, Kroupa01} does not influence the overall conclusions of this study. 

Turbulent star--forming gas and clump mass functions \citep{Reid05} indicate that a Salpeter--like IMF slope may be imprinted on the mass distribution of turbulent structures, and that all stellar clusters formed out of these structures therefore inherit a Salpeter--like IMF. The concept of a ``universal IMF" is, however, being increasingly scrutinised by recent studies based on large samples of galaxies and non--traditional approaches, all reporting discrepancies between the observations and model predictions. A number of recent studies now suggest an evolving or spatially varying IMF as a ``last resort'' explanation to reconcile the observed differences  \citep{HG08, W08b, Wilkins08a, V08, M09}. The assumed cosmological parameters are: $H_0=70$ km\,s$^{-1}$\,Mpc$^{-1}$, $\Omega_M = 0.3$ and $\Omega_\wedge = 0.7$.

\section[]{Observations}
\subsection{Galaxy And Mass Assembly survey}

Motivated by these recent failures of the universal IMF assumption, we have conducted an analysis exploring such variations using a sample of galaxies from the Galaxy And Mass Assembly (GAMA) survey \citep{Driver09, Robotham09, Baldry09}. 
GAMA is a spectroscopic survey with multi--wavelength photometric data undertaken at the Anglo--Australian Telescope using the 2dF fibre feed and AAOmega multi--object spectrograph. AAOmega provides 5\,\AA\, resolution spectra with complete spectral coverage from 3700--8800\,\AA\, \citep{Sharp06}. GAMA covers three equatorial fields of 48 deg$^2$ each, with two fields reaching a depth of $r_{AB}<19.4$ magnitude and the third extending to $r_{AB}<19.8$ magnitude, together with $K_{AB}<17.6$ magnitude over all three fields. There are $\sim120\,000$ galaxies with measured spectra available from GAMA observations to date \citep{Driver10}. The redshift of each galaxy is determined using RUNZ \citep{Saunders04}, a \textsc{fortran} program for measuring redshifts from reduced spectra. 

\subsection{Data}

The standard strong optical emission lines are measured from each curvature corrected and flux calibrated spectrum assuming a single Gaussian approximation and common values for redshift and line width (Bauer et al. 2011, in prep).  Corrections for the underlying stellar absorption, dust obscuration and fibre aperture effects, detailed below, are applied to these measurements. A full composite line and continuum extraction process is ultimately intended for the full data set.

The strength of H$\alpha$ emission in galaxy spectra is used in this investigation to probe the extent of the star formation in galaxies. The H$\alpha$ luminosity is used to measure the current SFR, as the ionising photons mainly come from short--lived massive stars. Our sample is drawn from the $\sim120\,000$ spectra available at June 2010, and is comprised of 43\,668 galaxies with measured emission lines, about 40\% of all galaxies in the GAMA sample at that time. This sample only includes objects with redshift quality flags $\geq$ 3 \citep[i.e.\,regarded as a secure redshift, see][]{Driver10}. Furthermore, we exclude all galaxies with H$\alpha$ emission measurements affected by the presence of strong sky lines (see Figure\,\ref{fig:BDvL}), and all galaxies with H$\alpha$ emission below a minimum flux limit of $25\times10^{-17} erg/s/cm^{2}$. This flux limit is obtained from examining the spectra of a sample of low H$\alpha$ luminosity galaxies. Increasing this flux limit to that used by \cite{Brough10} for a sample of low H$\alpha$ luminosity GAMA galaxies, for example, does not alter our conclusions. Sloan Digital Sky Survey (SDSS) photometry in {\em u,g,r,i,z\/} filters is available for each galaxy \citep{Hill10}. $k$-corrections to $z=0.1$ are applied and all photometry is corrected for foreground (Milky Way) dust-extinction \citep{Schlegel98}. The galaxy sample covers a moderate range in redshift ($0<z\leq0.35$). Galaxies dominated by emission from active galactic nuclei (AGN) are excluded from the sample ($5334$ galaxies) based on standard optical emission--line ([NII]/H$\alpha$ and [OIII]/H$\beta$) diagnostics using the discrimination line of \cite{Kewley01}.  In the case of galaxies for which only some of these four emission lines are measurable, AGNs can still be excluded using the diagnostics $\log$ ([NII]/H$\alpha$)$\geq$0.2 and $\log$ ([OIII]/H$\beta$)$\geq$1. This excludes a further 173 galaxies. The size of the final sample is $33\,657$.
\begin{figure}
\begin{center}
\includegraphics[scale=0.48]{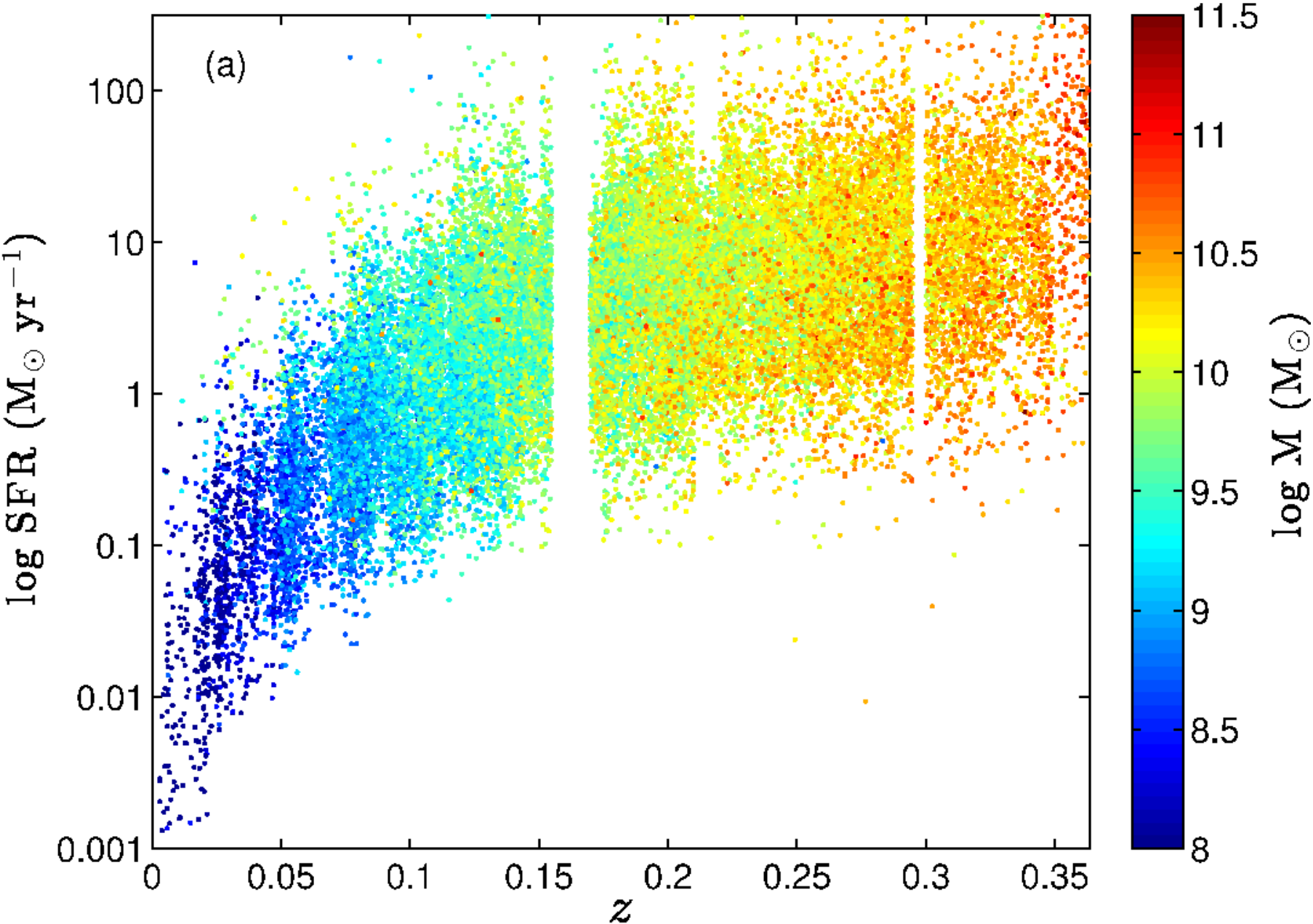}
\includegraphics[scale=1]{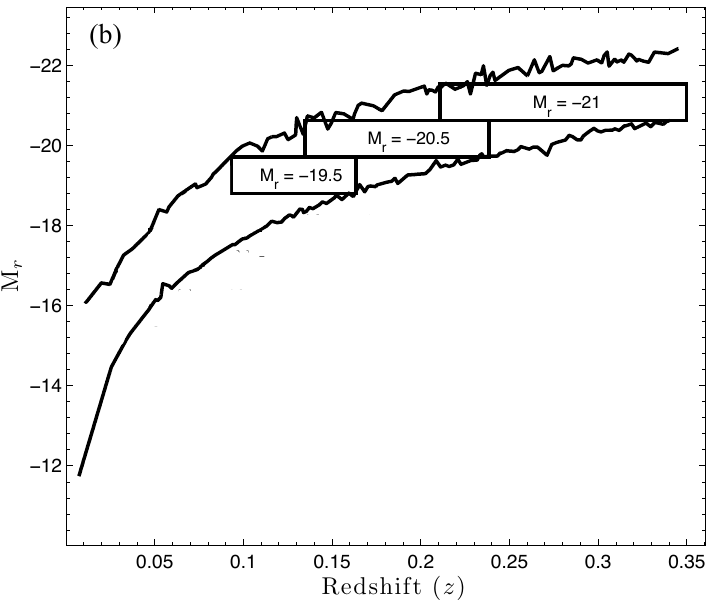}
\caption{{(a) The distribution of SFR and galaxy stellar masses with redshift. The visible gap in the distribution centred on $z=0.16$ shows where the wavelength of the atmospheric O$_2$ band (Fraunhofer A--line) overlaps with the redshifted wavelength of the H$\alpha$ emission line, leading us to omit these data from our analysis. The masses are in units of the solar mass, 1 M$_{\odot} = $ $1.99\times10^{30}$ kg. (b) The definition of the volume limited samples used in this study. The galaxies within each magnitude range are not affected by the flux limits of the survey, so that each of the three samples is complete to a given luminosity.}}
\label{fig:sfr_z_mass}
\end{center}
\end{figure}

 This sample of galaxies spans a large range in stellar mass ($7\leq \log\, (M/M_{\odot}) \leq 12$) and SFR ($10^{-3}$--100 M$_{\odot}$yr$^{-1}$). It is this large range in SFR, stellar mass and redshift that permits us to explore the potential IMF dependencies with respect to different physical properties of galaxies. Figure~\ref{fig:sfr_z_mass}(a) shows the wide range in SFR  sampled as a function of redshift, and colour coded to illustrate the range in stellar masses.  Figure\,\ref{fig:sfr_z_mass}(b) shows the envelope of the distribution of absolute $r$--band magnitude, $M_r$, with redshift for this sample. Outlined within this envelope are three independent volume--limited samples which form the basis of our subsequent analysis. These are selected to span $\sim1$ magnitude in $M_r$, centred on the values shown.

\section{Deriving physical quantities}

\subsection{H$\alpha$ luminosities}

The observed H$\alpha$ \, emission must be corrected for the effects of stellar absorption within the host galaxy, obscuration by dust, and the limited sampling of each galaxy by the optical fibres (aperture effects) used in multi--object spectroscopy. 

\subsubsection{Stellar absorption correction and Balmer decrements}
A simple constant correction for stellar absorption in Balmer emission line equivalent widths (EWs) (i.e.\,H$\alpha$ and H$\beta$ EWs) is used for this investigation. The assumed common EW correction (EW$_c$) for the stellar absorption in the GAMA data is $1.3$\,\AA. Based on previous work \citep{Hopkins03}, a correction of at most $1.3$\,\AA \, is sufficient, provided the assumption is restricted to studies examining the gross characteristics of a large sample of sources, which is the case in this investigation. We tested a range of EW$_c$ values between $0.7 $ and $1.3$\,\AA, and the results did not vary measurably.  Only H$\alpha$ EWs smaller than $\log$ (H$\alpha$ EW)$<0.9$ show a difference in EW of more than $5\%$.

As the stellar absorption may in general be a luminosity dependent quantity, the required correction could be higher for high star formation rate sources. In order to test this aspect, a unique EW$_c$ for each galaxy is assigned based on an assumed linear relationship between luminosity and stellar absorption correction. The results indicate that a luminosity dependent stellar absorption correction does not significantly affect the calculated intrinsic luminosity of the source. Less than $10\%$ of the sample showed any noticeable effect.  For systems with $\log$ (H$\alpha$ EW)$>2$, the assumption of a luminosity dependent absorption correction increases the inferred EWs for these extreme systems, thereby further enhancing the trend with SFR presented in this paper. Therefore, the assumption of a fixed EW$_c=1.3$\,\AA\, should not significantly affect the trends evident in the results, or our conclusions. 

 The Balmer Decrement (BD) is defined as the ratio of stellar absorption corrected H$\alpha$ to H$\beta$ fluxes ($BD= S_{H\alpha}/S_{H\beta}$), where $S_{H\alpha}$ for example is \citep{Hopkins03},
\begin{equation}
S_{H\alpha} = F_{H\alpha} \times \frac{(H\alpha EW + EW_c)}{H\alpha EW},
\end{equation}
where $F_{H\alpha}$ is the measured emission line flux.
\begin{figure}
\begin{center}
\includegraphics[scale=0.55]{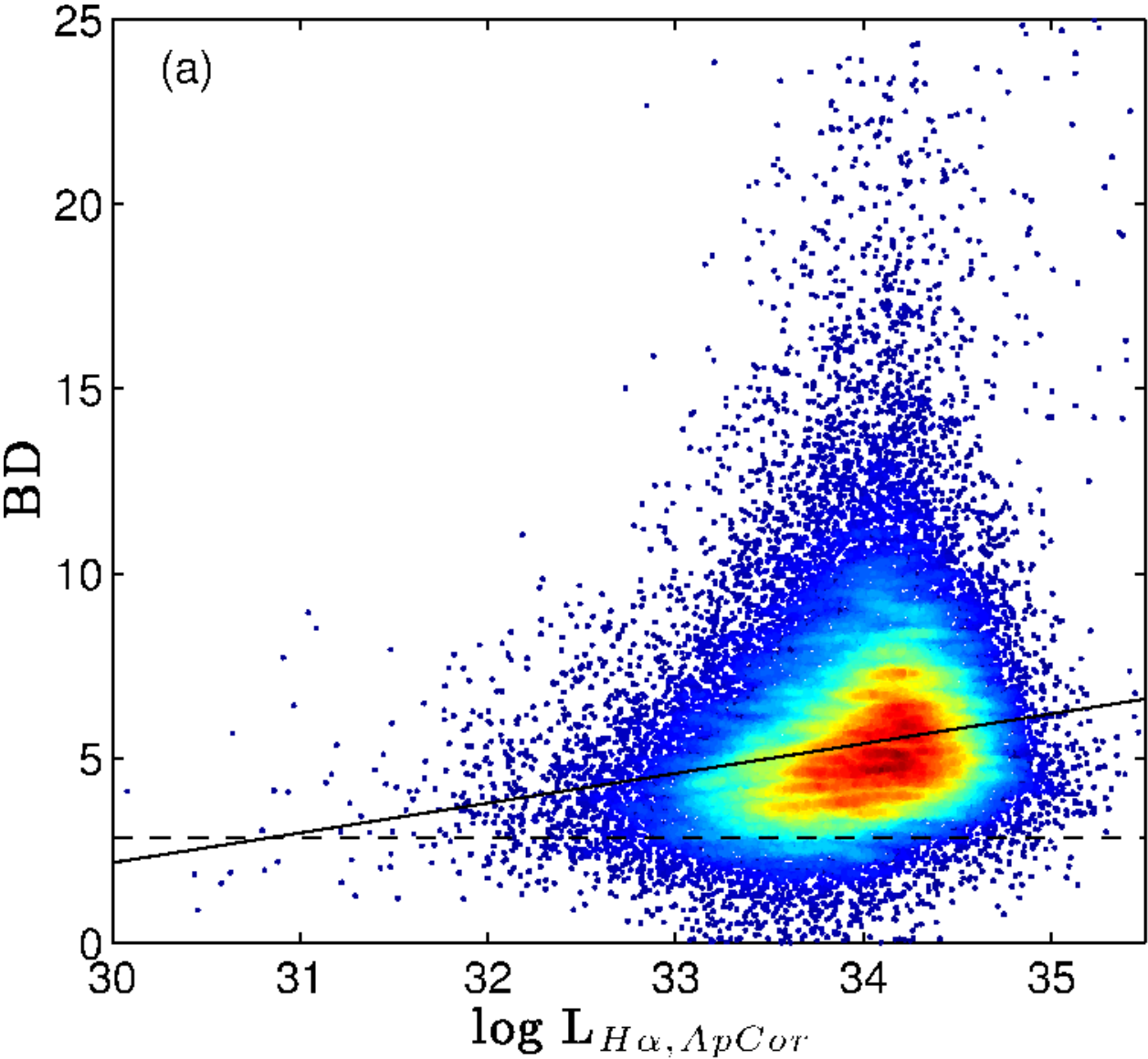}
\includegraphics[scale=0.48]{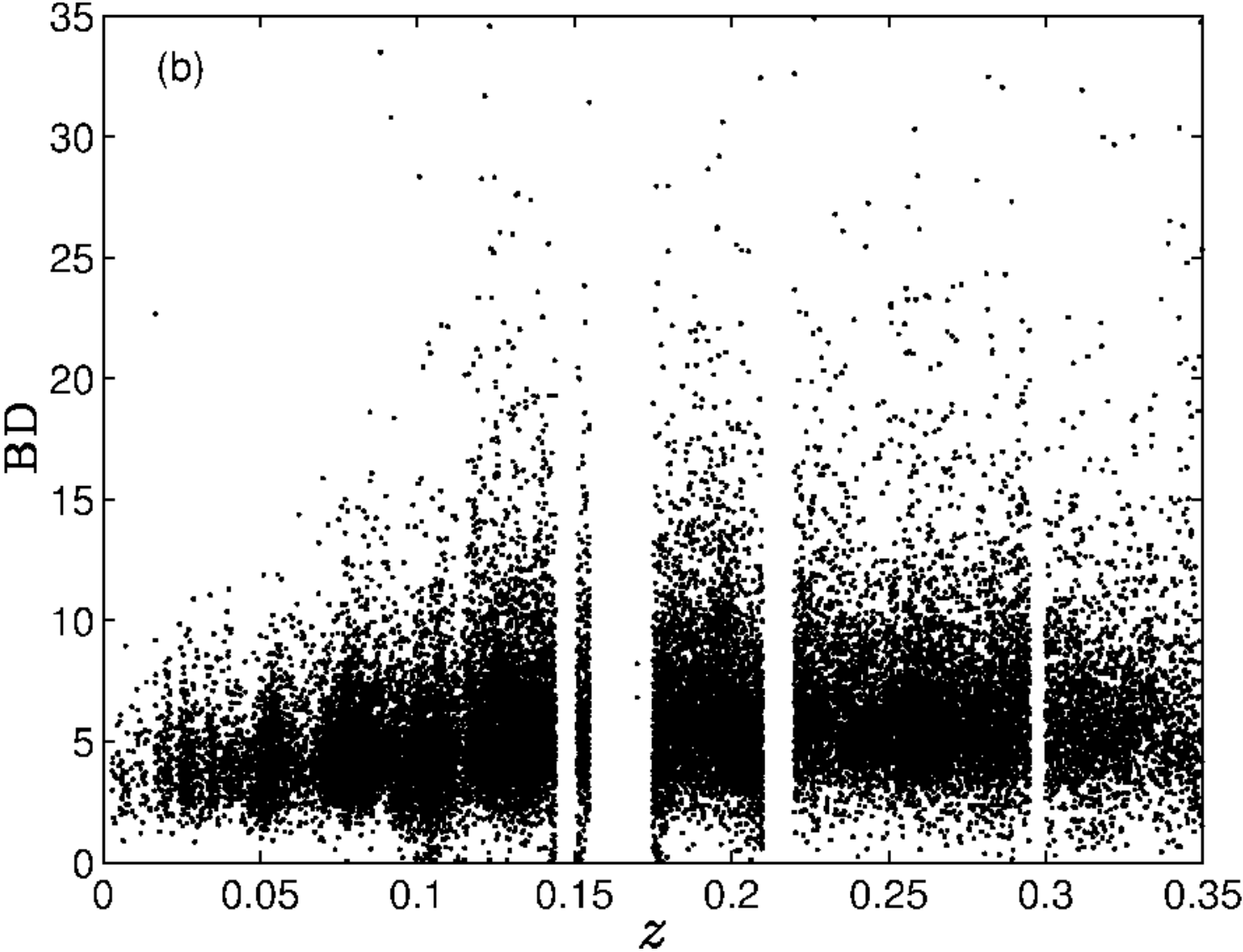}
\caption{ (a) The relationship between Balmer Decrements and aperture corrected luminosities for the galaxies with accurate H$\alpha$ and H$\beta$ EW measurements. A linear relation is fitted to the data to determine the Balmer Decrements for the galaxies that have measured H$\alpha$ EWs but no H$\beta$ EWs or have H$\beta$ EWs that are affected by sky absorption bands. (b) The dust obscuration as indicated by the Balmer Decrement is largest at high $z$. This is because high SFR objects also tend to have higher obscurations \citep{Hopkins01} and the objects at higher $z$ tend to have higher SFRs, a result of both galaxy evolution and a predominance of massive, high--SFR galaxies at high--$z$, due to the flux limit of the survey.}
\label{fig:BDvL}
\end{center}
\end{figure}

A small fraction of galaxies have Balmer Decrements (BD) less than the Case B recombination value BD$=2.86$ (Figure~\ref{fig:BDvL}). The Balmer Decrement is an obscuration sensitive parameter, and its departure from the Case B recombination value of 2.86 is an indication of the dust attenuation along the line of sight. BD$<2.86$ can result from an intrinsically low reddening combined with uncertainty in the stellar absorption, but also from errors in the line flux calibration and measurements \citep{Kewley06}. Out of all the galaxies with measured Balmer Decrements, $\sim 6\%$ have BD$<2.86$. All these Balmer Decrements are set to 2.86 for the purpose of this investigation.

Not all galaxies have both H$\alpha$ and H$\beta$ measurements. For the galaxies with only H$\alpha$ measurements, the relation between aperture corrected luminosity and Balmer Decrement is used to determine the Balmer Decrements. The empirical form of the relationship between aperture corrected luminosity and Balmer Decrement is shown in Figure~\ref{fig:BDvL}. The form is
\begin{eqnarray}
BD &=& 0.81\log(L_{H\alpha, ACor}) - 22.041,\ \ \mbox{$\log(L_{H\alpha, ACor})>30.74$} \nonumber \\
&=& 2.86, \ \  \mbox{$\log(L_{H\alpha, ACor})\leq30.74$}
\label{eqn:BD}
\end{eqnarray}
where $\log(L_{H\alpha, ACor})$ denotes the aperture corrected H$\alpha$ luminosity (see Eqn.~\ref{maths:apertureCor_lum}).

\subsubsection{Aperture correction}

\begin{figure}
\begin{center}
\includegraphics[scale=0.5]{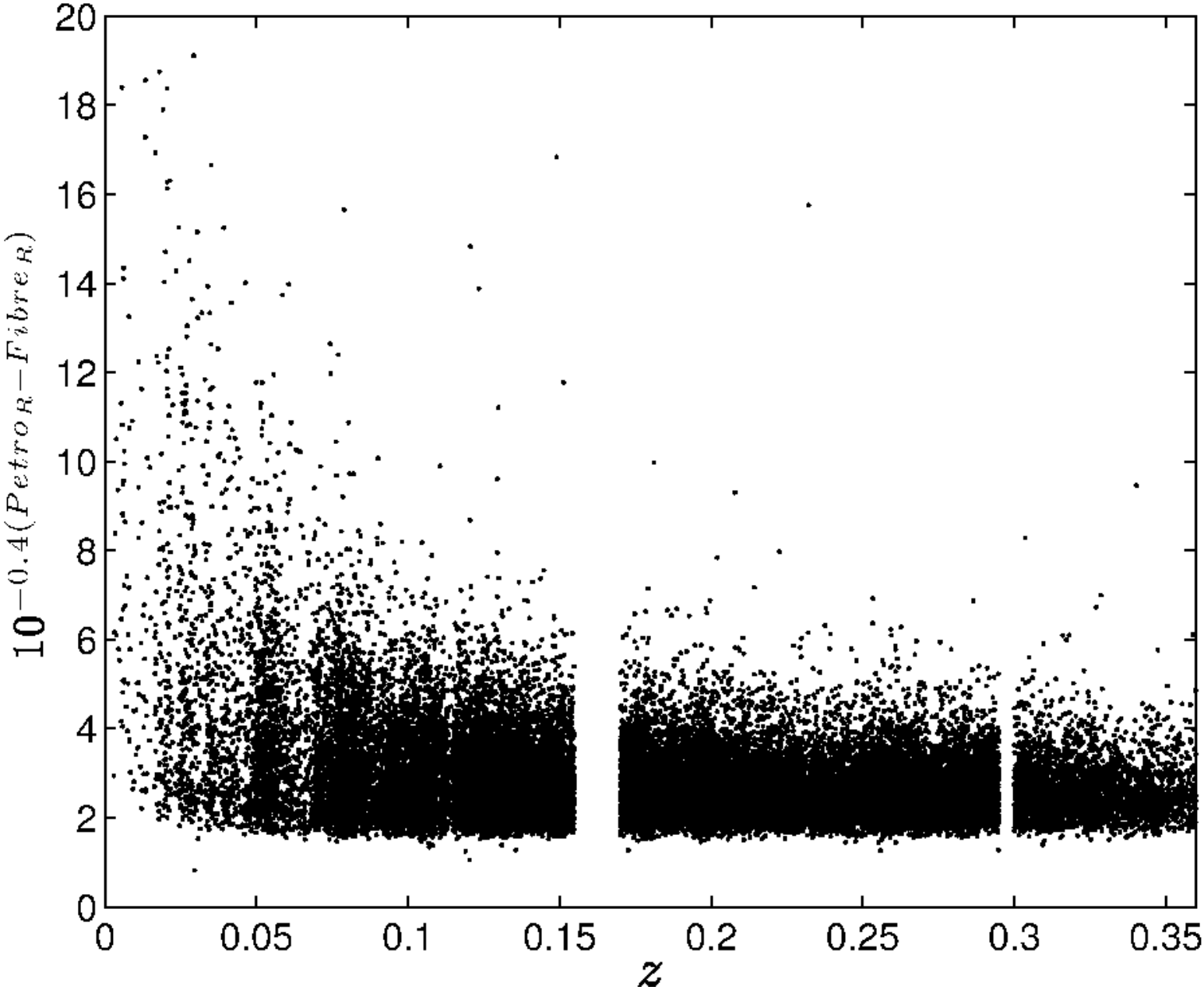}
\caption{ Aperture correction factor as a function of redshift. The required aperture correction is largest at low $z$ because low $z$ objects are more likely to be larger in angular size than the aperture of the spectroscopic fibre used for the observations.}\label{fig:BDvz}
\end{center}
\end{figure}

Aperture effects arise from the physical limitation imposed by the diameter of the spectroscopic fibre used in the observations. For nearby sources, this means that the fibre only captures part of the light from the object, which is naturally a problem for sources larger in size than the fibre diameter projected on the sky. An aperture correction is required to account for the missing flux, in order to get an estimate of the true star formation rate. Following the approach used for SDSS spectra by \citet{Hopkins03}, we implemented an aperture correction for the H$\alpha$ luminosities of the GAMA galaxies. This approach uses the absolute $r$--band magnitude to approximate the continuum at the wavelength of H$\alpha$, thereby accounting for the H$\alpha$\, luminosity of the whole galaxy (Eq.\,\ref{maths:apertureCor_lum}). Figure\,\ref{fig:BDvz} shows the relation between the applied aperture correction and $z$ and the required correction is typically a factor of $2-4$.

The aperture corrected H$\alpha$ luminosity ($L_{H\alpha, ACor}$) for the whole galaxy is a function of three parameters, H$\alpha$ EW (EW$_{H\alpha, obs}$), absolute $r$--band luminosity ($M_r$) and redshift ($z$).  The form of the aperture corrected luminosity before applying any obscuration correction is \citep{Hopkins03}

\begin{eqnarray}
L_{H\alpha, ApCor} &= &(EW_{H\alpha, obs} + EW_c) \times 10^{-0.4(M_r - 34.10)} \nonumber \\ 
&& \times \frac {3\times 10^{18}}{(6564.61(1+z))^2}.
\label{maths:apertureCor_lum}
\end{eqnarray}

 The aperture corrections are based on the absolute magnitude, $M_r$, of each galaxy  as an estimate of  continuum luminosity, thereby recovering a H$\alpha$ luminosity for the whole galaxy. This method of  applying aperture corrections to the luminosities, described in \cite{Hopkins03}, yields similar results to the more complex colour gradient based aperture corrections described in \cite{BCW04}. We use these aperture corrected values throughout this analysis. We have, in addition, tested the effect on our results in the case of using SFRs estimated only from the detected H$\alpha$ emission through the fibre (no aperture corrections). Even in this case, we find the same qualitative conclusions regarding the SFR--dependence of IMF slope. We conclude that the aperture corrections applied here are not introducing any significant bias, nor are they erroneously giving rise to our results.

\begin{table}
\caption{The SFR calibration factors for different IMFs}
\begin{center}
\begin{tabular}{|c|c|}
\hline
high--mass slope  &  calibration factor (W)\\
\hline
$\alpha=-2.00$     & $4.08\times10^{34}$\\
$\alpha=-2.35$     & $1.27\times10^{34}$\\
$\alpha=-2.50$     & $0.65\times10^{34}$\\
\hline
\end{tabular}
\label{table1}
\end{center}
\end{table}

The aperture, obscuration and stellar absorption corrected luminosity for the whole galaxy is given as
\begin{eqnarray}
L_{H\alpha, int} &=& (EW_{H\alpha} + EW_c) \times 10^{-0.4(M_r - 34.10)} \nonumber \\
&& \times \frac {3\times 10^{18}}{(6564.61(1+z))^2} \Big (\frac{F_{H\alpha}/F_{H\beta}}{2.86} \Big)^{2.36},
\label{maths:apertureObsCor_lum}
\end{eqnarray}
where $F_{H\alpha}/F_{H\beta}$ denotes the Balmer decrement. Figure\,\ref{fig:BDvL}(b) explores the increase in Balmer decrement with respect to increasing $z$. The exponent of the Balmer Decrement in Eq.\,\ref{maths:apertureObsCor_lum} is equal to $k(\lambda_{H\alpha})/[k(\lambda_{H\beta}) - k(\lambda_{H\alpha})]$, where extinction at a given $\lambda$, $k(\lambda)$, is determined from the \cite{Cardelli89} galactic dust obscuration curve.

The H$\alpha$ SFR can be determined from \cite{Kennicutt98} 

\begin{equation}
SFR_{H\alpha} = \frac{L_{H\alpha}}{1.27\times 10^{34}}.
\label{maths:SFR}
\end{equation}

The derivation of the H$\alpha$ SFR calibration requires the assumption of an IMF. The above calibration factor has been derived assuming a Salpeter IMF. 
Table\,\ref{table1} shows the effect on the SFR calibrator if a different IMF is assumed. For a given luminosity, a calibration based on a flatter IMF would indicate a lower SFR than the Salpeter IMF based calibration.

If an IMF dependent SFR calibration is used in the derivation of SFRs for GAMA galaxies, that would reduce the range in SFR shown in Figure 1(a). This reduction in range would not affect our main conclusion of an IMF--SFR relationship, because the SFR calculated for the sample would still vary monotonically (as the scaling is linear, the ordering of the SFRs is not affected).

\subsection{The determination of stellar masses}

The stellar masses used in this investigation are derived based on the observed tight relation between ($g-i$) colour and the mass--luminosity ($M/L$) relation \citep[ 2011; in prep]{Ned09} using the \cite{BC03} models and assuming a \cite{Chabrier03} IMF. This method of calculating the masses yields results consistent with other established techniques \citep[e.g.,][]{Baldry06}.
\begin{equation}
\log (M_*) = -0.68 + 0.73(g - i) - 0.4(M_i - M_{i,\odot}),
\label{maths:masses}
\end{equation}

\noindent where $M_*$ is the mass of the galaxy, $g$ and $i$ band colours are $k$-corrected to $ z=0$, $M_i$ is the absolute magnitude of the galaxy in the $i$-band  and $M_{i, \odot} = 4.58$, the absolute magnitude of the Sun in the $i$-band.

\section{Obscuration corrections}

Understanding and interpreting the physical and chemical properties of galaxies depends in part on how accurately the data are corrected for stellar absorption and dust obscuration to recover the intrinsic fluxes. H$\alpha$ can be heavily attenuated by dust. High SFR galaxies are subjected to greater dust obscuration than lower luminosity objects \citep{Hopkins01, Hopkins03, Afonso03, Perez03}. Obscuration corrections are especially critical in this analysis as our primary aim is to compare the observed H$\alpha$ EW and $g-r$ (or $g-i$) colours with P\'{E}GASE generated synthetic spectra for different input IMFs, assuming no extinction. 

The reliability of the applied dust correction depends on the adopted dust obscuration models. We explore several popular empirical dust models \citep{Cardelli89, Calzetti01, FD05} along with radiative transfer model predictions of the effects of dust extinction \citep{Popescu00, Tuffs04, Popescu11}.

The differential reddening between the stellar continuum and gas \citep{Calzetti01} must be addressed in deriving the intrinsic fluxes.  The difference in attenuation between gas and continuum is generally assumed to be $\sim 2$ \citep{WAS10, M09, HG08, Calzetti01}. \cite{HG08} describe the effect, on H$\alpha$ EWs and colours, of varying the differential reddening factor between the continuum and gas. We tested the impact of this assumption on our results. For all the subsequent analysis in this paper we use obscuration corrected colours derived through the application of an obscuration curve together with this factor of $\sim2$ (as detailed below). These measurements were compared against colours derived by Taylor et al. (in prep), from full SED modelling of the GAMA photometry, with independent dust corrections. The results are consistent, with an RMS scatter of $\sim0.17$ mag, with no systematic deviation, as might be expected if the factor of $\sim2$ between gas and continuum obscuration were significantly in error. There is, moreover, no systematic offset in these two approaches as a function of SFR, specific SFR, mass or redshift. We conclude that our assumption of this commonly used factor is justified, and unlikely to introduce any systematic error in our result. 

The obscuration corrected H$\alpha$ EW is given as 

\begin{eqnarray}
EW_{H\alpha, int}&=& \frac {(EW_{H\alpha, obs}+1.3)}{(1+z)}\nonumber \\
&& \times10^{0.4[k(\lambda_{H\alpha})E(B-V)_{gas} - k(\lambda_{H\alpha})E(B-V)_{*}]}, 
\label{maths:HaEW}
\end{eqnarray}
where  k$(\lambda)$ gives the extinction at wavelength $\lambda$, and the colour excess of gas (i.e., emission lines) is

\begin{equation}
E(B-V)_{gas} = \frac {\log \Big ( \frac{H\alpha}{H\beta}/2.86 \Big )}{0.4[k(\lambda_{H\beta}) - k(\lambda_{H\alpha})]}.
\end{equation}

The colour excess of the continuum \citep{Calzetti97} is
\begin{equation}
E(B-V)_{*} = 0.44E(B-V)_{gas}.
\end{equation}

The effect of the spectroscopic fibre sampling only the central regions, for those galaxies largest on the sky, may be to limit the detection of the lowest EW systems in low SFR galaxies. Accounting for this effect would not change our results; indeed, if anything, such a correction would act to enhance the trend investigated here.

\subsection{Applying obscuration corrections}

This section describes the different obscuration curves used in this analysis, and the methods of applying obscuration corrections to the data. 

We have used a combination of  the \citet{Calzetti01} and \cite{Cardelli89} obscuration curves and the \cite{FD05} curve as described by \cite{WAS10} to determine the necessary corrections. These extinction curves and the application of dust corrections are described below. 

\subsubsection{\bf \citet{Calzetti01} obscuration law}

This dust extinction curve is appropriate for continuum attenuation corrections as this curve is derived from spatially integrated colours of the entire stellar population in a sample of starburst galaxies.  Embedded within the analytical form of this curve are dust geometry and composition, and it mostly describes dust absorption, since the effects due to scattering are averaged out. Because the entire stellar population within the galaxies was observed, emission lost through dust scattering out of the line of sight is averaged out by the scattering into the line of sight. The form of the curve is

\begin{eqnarray}
k(\lambda) & = & 2.656 \Big ( -1.857 + \frac {1.040}{\lambda} \Big ) + 4.05, \nonumber \\ 
&& \mbox {for $0.63\mu$m $\leq \lambda \leq 2.2\mu$m} 
\label{maths:calzettiCurve}
\nonumber
\\
\nonumber
& = & 2.656 \Big ( -2.156 + \frac {1.509}{\lambda} - \frac{0.198}{\lambda^2} + \frac{0.011}{\lambda^3} \Big ) + 4.05, \nonumber \\ 
&& \mbox{for $0.12\mu$m $\leq \lambda < 0.63\mu$m}.
\end{eqnarray}

The second term of the exponent in Eq.\,\ref{maths:HaEW}, which is related to the continuum luminosity, and the corrections to the colours are derived using this curve.

\subsubsection{\bf \citet{Cardelli89} obscuration law}

This Galactic dust obscuration curve is derived from observations of the UV extinction of stars, as well as using various other sources for optical and NIR data. The young stellar populations in massive star forming regions responsible for UV radiation are also responsible for nebular emission lines and this curve is applicable to both diffuse and dense stellar regions. This curve accurately describes the dust effects on emission lines and has the form

\begin{equation}
k(\lambda) = a(x) + \frac {b(x)}{R_v}, 
\label{maths:cardelli_1}
\end{equation}

\noindent where $R_v$ is the ratio of total to selective extinction and is a constant for a given extinction curve. The value of $R_v=3.1$ \citep{Calzetti01}, which is found to well describe the reddening of the ionised gas in star forming galaxies, is used in this analysis.

The functional forms of $a(x)$ and $b(x)$ in Eq.\,\ref{maths:cardelli_1}, with $x=1/\lambda$, are a power law in the infrared regime and a polynomial in the optical/NIR regime, in units of $\mu$m$^{-1}$:
\\

\noindent Infrared:  $0.3\mu$m$^{-1}$ $\leq x \leq 1.1\mu$m$^{-1}$;
\begin{eqnarray}
a(x)  &=&  0.574 \lambda^{1.61}; \nonumber \\
b(x) &=&  -0.527 \lambda^{1.61}. \nonumber
\end{eqnarray}
Optical/NIR:  $1.1\mu$m$^{-1}$ $\leq x \leq 3.3\mu$m$^{-1}$ and $y = (x-1.82)$;
\begin{eqnarray}
a(x)  &=& 1+0.17699y-0.50447y^2-0.02427y^3+0.72085y^4 \nonumber \\
&& +0.01979y^5-0.7753y^6+0.32999y^7; \nonumber \\
b(x) &=& 1.41338y+2.28305y^2+1.07233y^3-5.38434y^4 \nonumber \\
&& -0.62251y^5+5.30260y^6-2.09002y^7.
\end{eqnarray}

The first term of the exponent in Eq.\,\ref{maths:HaEW}  related to the line luminosity is based on this curve. The intrinsic H$\alpha$ EW is the ratio of the corrected H$\alpha$ line to continuum luminosities.

\subsubsection{\bf \citet{FD05} obscuration law}

A recent study by \cite{WAS10}, looking at dust obscuration in galaxies using GAMA data, tested a number of common obscuration curves, including the \cite{Calzetti01, Calzetti97}, and \cite{Cardelli89} dust curves. They found that a \cite{FD05} obscuration curve with R$_v=4.5$ and the $2200$\AA\,bump removed gives an excellent agreement between far ultraviolet, near ultraviolet, H$\alpha$ and [OII] derived star formation rate indicators.

In this case, both terms of the exponent in Eq.\,\ref{maths:HaEW} and the corrections to the colours are determined using the \cite{FD05} curve.

\subsection{\bf \citet{Popescu00, Popescu11} and \citet{Tuffs04} radiative transfer models}

In addition to the dust corrections based on the above dust obscuration curves, the effects of dust attenuation on H$\alpha$ EW and $g-r$ parameters can be determined using radiative transfer models \citep{Popescu00,Tuffs04, Popescu11}, where the attenuation of star light from disk galaxies with different dust geometries of different stellar ages constrained by UV/optical to FIR/submm spectral energy distributions are considered. The model predictions are based on the opacity of the diffuse dust component, given as face--on B--band optical depth ($\tau_b$), inclination of a galaxy and a clumpiness factor (F) describing the local absorption of UV light from massive stars due to the presence of massive star forming regions. 

\begin{figure*}
\begin{center}
\includegraphics[scale=0.52]{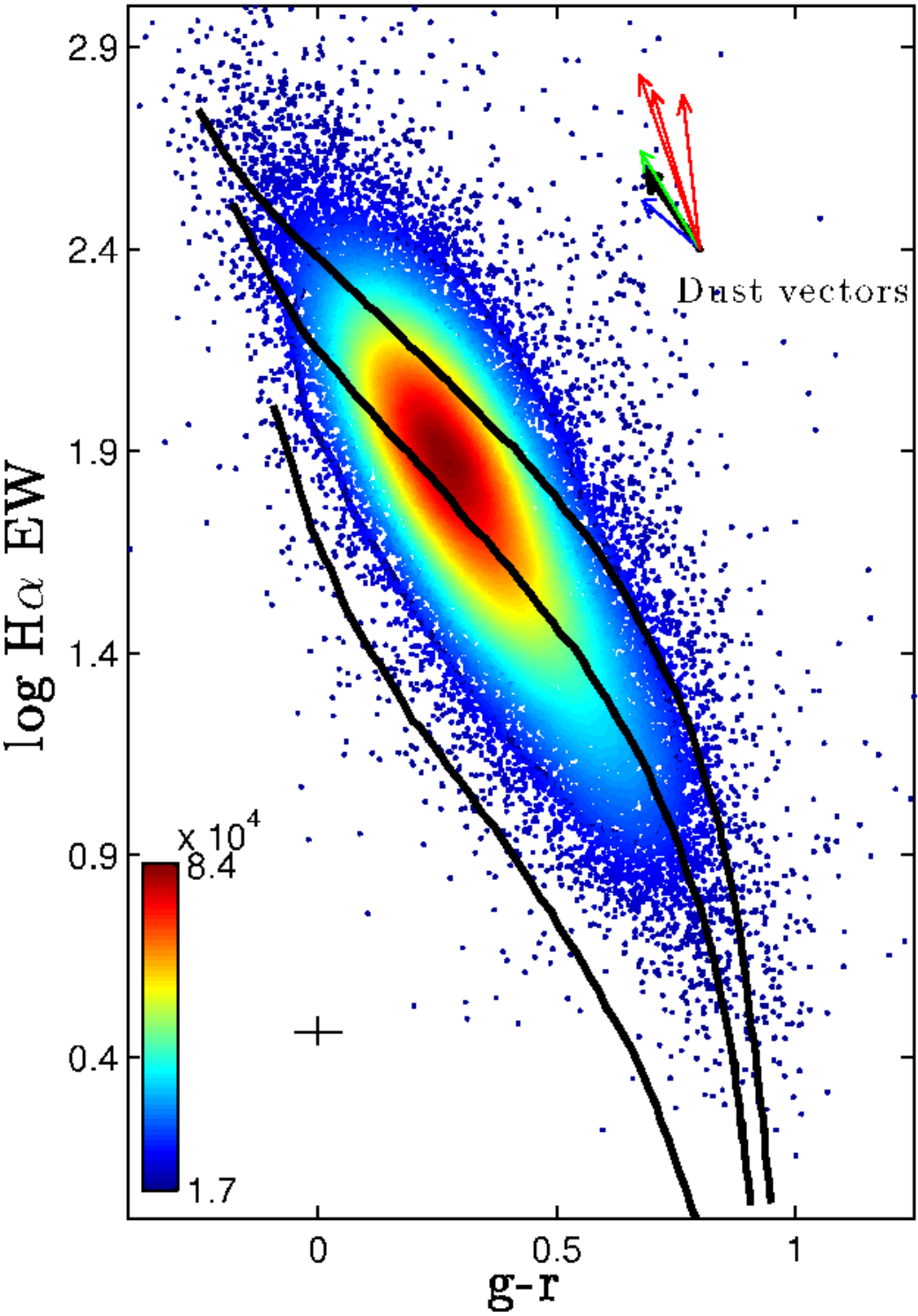}
\caption{Distribution of all GAMA galaxies up to z=0.355, after dust corrections as given in the text.  All data and the model tracks are $k$-corrected to z=0.1. The colour contours indicate the data density and the three solid lines indicate the three different evolutionary paths a galaxy would take if all star clusters within that galaxy have an IMF with a slope of $\alpha=-3$ (bottom track), $\alpha=-2.35$ (middle track) or $\alpha=-2$ (top track). These model tracks are generated using P\'EGASE. The arrows depict the dust vectors. The red arrows represent radiative transfer model predictions calculated using the model of \citet{Popescu00, Popescu11} and \citet{Tuffs04} and from left to right correspond to $\tau_b=8, 4, 1$, all assuming a median galaxy inclination of $60^{\circ}$ and $F=0.35$. The rest of the vectors show the movement of data points for different dust extinction curves and for a Balmer Decrement of $4$. {\it Blue}: The dust vector calculated using the \citet{Calzetti97} curve for the continuum corrections and \citet{Cardelli89} curve for emission line corrections. {\it Green}: The dust vector corresponding to corrections calculated using \citet{FD05} curve as modified by \citet{WAS10}. {\it Black}: The dust vector corresponding to the \citet{Calzetti01} and \citet{Cardelli89} curves for the continuum and emission corrections respectively.} 
\label{fig:complete_sample}
\end{center}
\end{figure*}

These various approaches to obscuration corrections are detailed as vectors in Figure\,\ref{fig:complete_sample}, showing the effect of different obscuration curves, or models, on the data, for an assumed Balmer decrement of $BD=4$. As shown in more detail below, the different approaches to dust correction do not change our qualitative conclusions.

\section{Evidence for a non--universal IMF}\label{sec:SFR}

\begin{figure*}
\begin{center}
\includegraphics[scale=0.65]{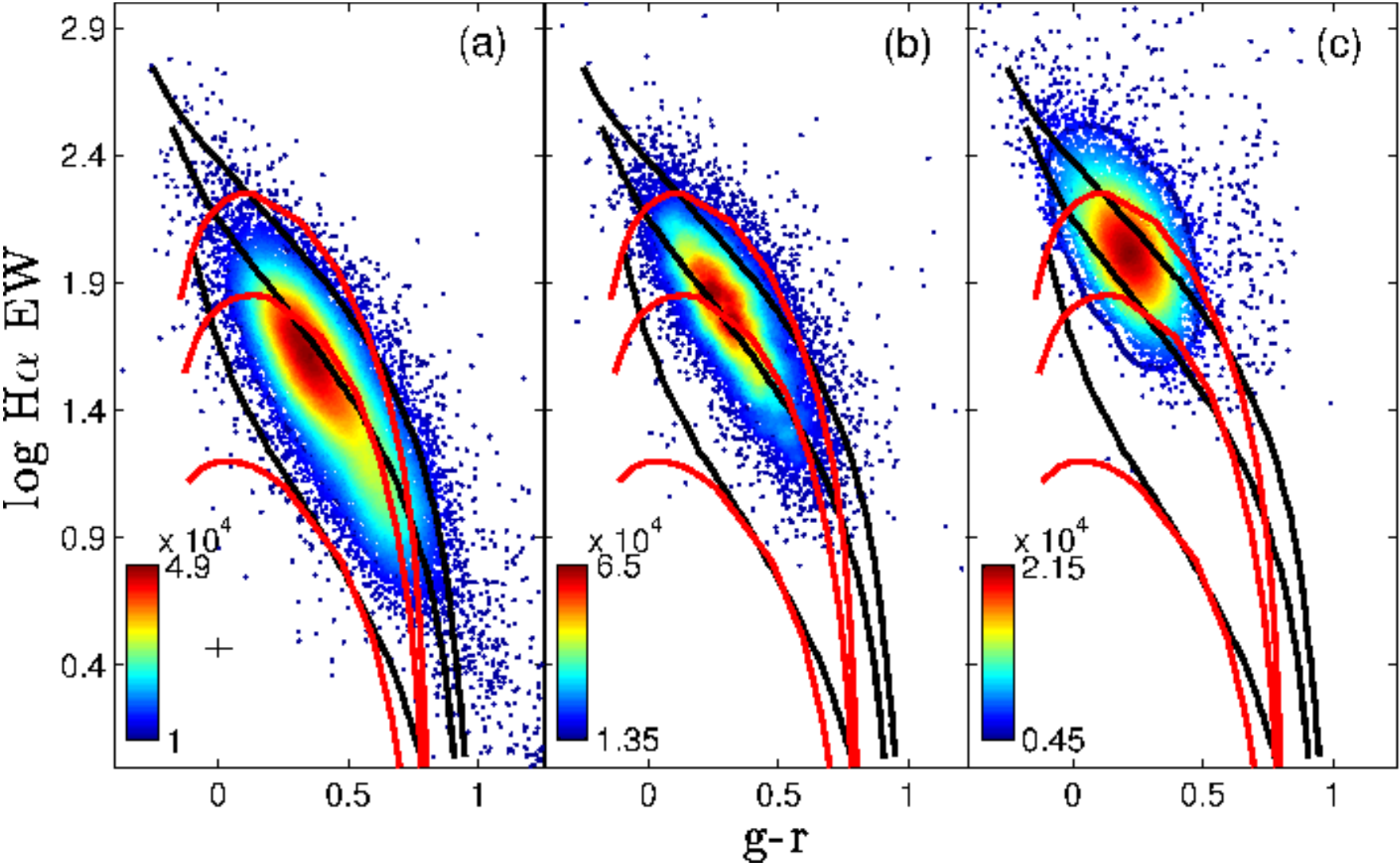}
\caption{The sample of GAMA galaxies divided into three sub--samples based on SFRs. (a) $0<$SFR (M$_{\odot}$yr$^{-1}$) $<3$, (b) $3\leq $SFR (M$_{\odot}$yr$^{-1}$) $<13$ and (c) SFR (M$_{\odot}$yr$^{-1}$)$\geq 13$. The two sets (black and red) of three solid lines indicate the three different evolutionary paths a galaxy would take in H$\alpha$\, EW and $g-r$ colour if all star clusters within that galaxy have an IMF with a slope of $\alpha=-3$ (bottom track), $\alpha=-2.35$ (middle track) or $\alpha=-2$ (top track). The black lines are the evolutionary paths predicted by P\'EGASE \citep{FR97} and red lines are paths predicted by \citet{M05} models. The age increases along the tracks from 100 Myr (top left) to 13 Gyr (bottom right). Coloured contours are drawn based on data densities of each sub--sample. The ranges in data densities are indicated alongside the colour bars of each plot. A representative uncertainty on individual measurements is indicated by the error bars in the bottom left of (a). A variation with SFR is apparent across the three panels, with high star forming sources evidently preferring a flatter IMF.}
\label{fig:sfrBins}
\end{center}
\end{figure*}

The three model evolutionary tracks shown as black lines in Figure~\ref{fig:sfrBins} are reproduced from \cite{HG08}. These evolutionary tracks are generated using the population synthesis code P\'EGASE \citep{FR97}, under the assumption of no extinction and an exponentially declining star formation history with an $e$--folding time of $1.1$\,Gyr. The model evolutionary tracks denoted by the red lines are generated by combining \cite{M05} and P\'EGASE models. Based on a similar analysis of H$\alpha$\, EW and $g-r$ colour, \cite{HG08} suggest a possible systematic variation in the IMF slope, in which faint galaxies prefer steep IMFs. Our results, (Figure~\ref{fig:sfrBins}) shown for three sub--samples based on SFR, are consistent with those of \cite{HG08}, and are not sensitive to the choice of population synthesis models. The low luminosity systems, which also have low SFRs (Figure~\ref{fig:sfrBins}(a)), lie below the central model track representative of a Salpeter IMF ($\alpha=-2.35$) and towards the bottom track with $\alpha=-3$. In contrast, those with high SFRs lie towards the top model track with $\alpha=-2$. 
\begin{figure*}
\begin{center}
\includegraphics[scale=0.44]{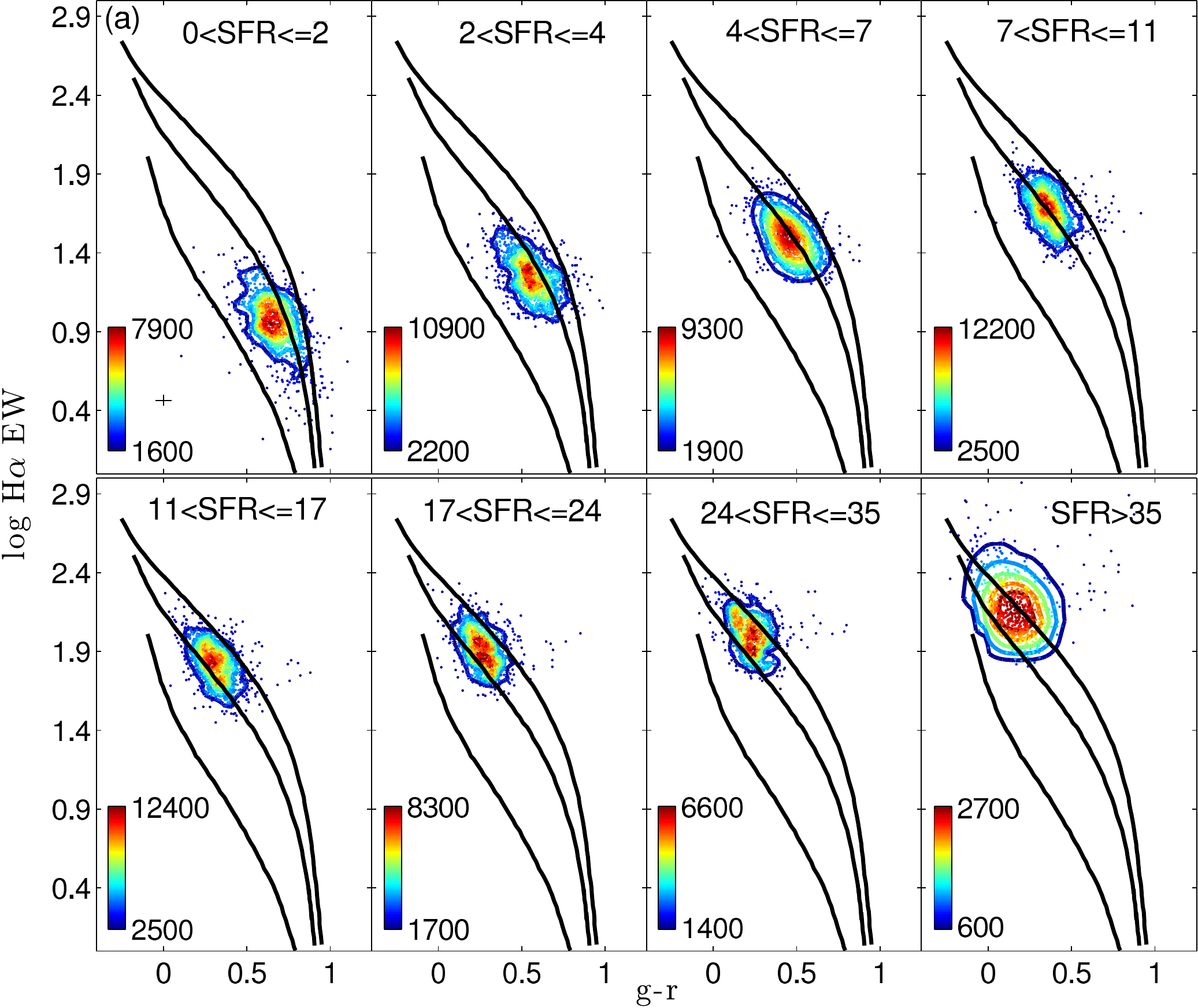}\\
\includegraphics[scale=0.44]{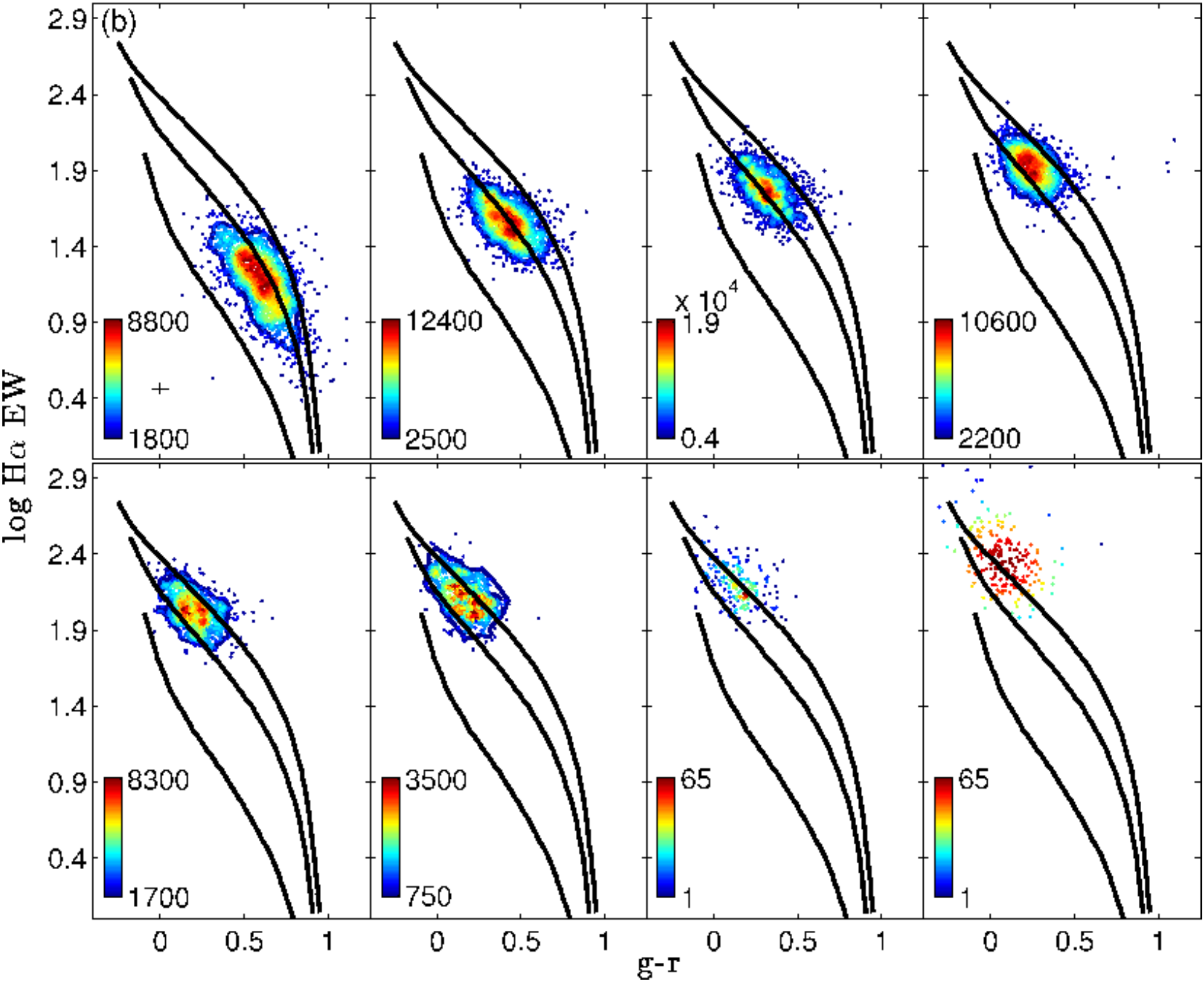}\\
\includegraphics[scale=0.44]{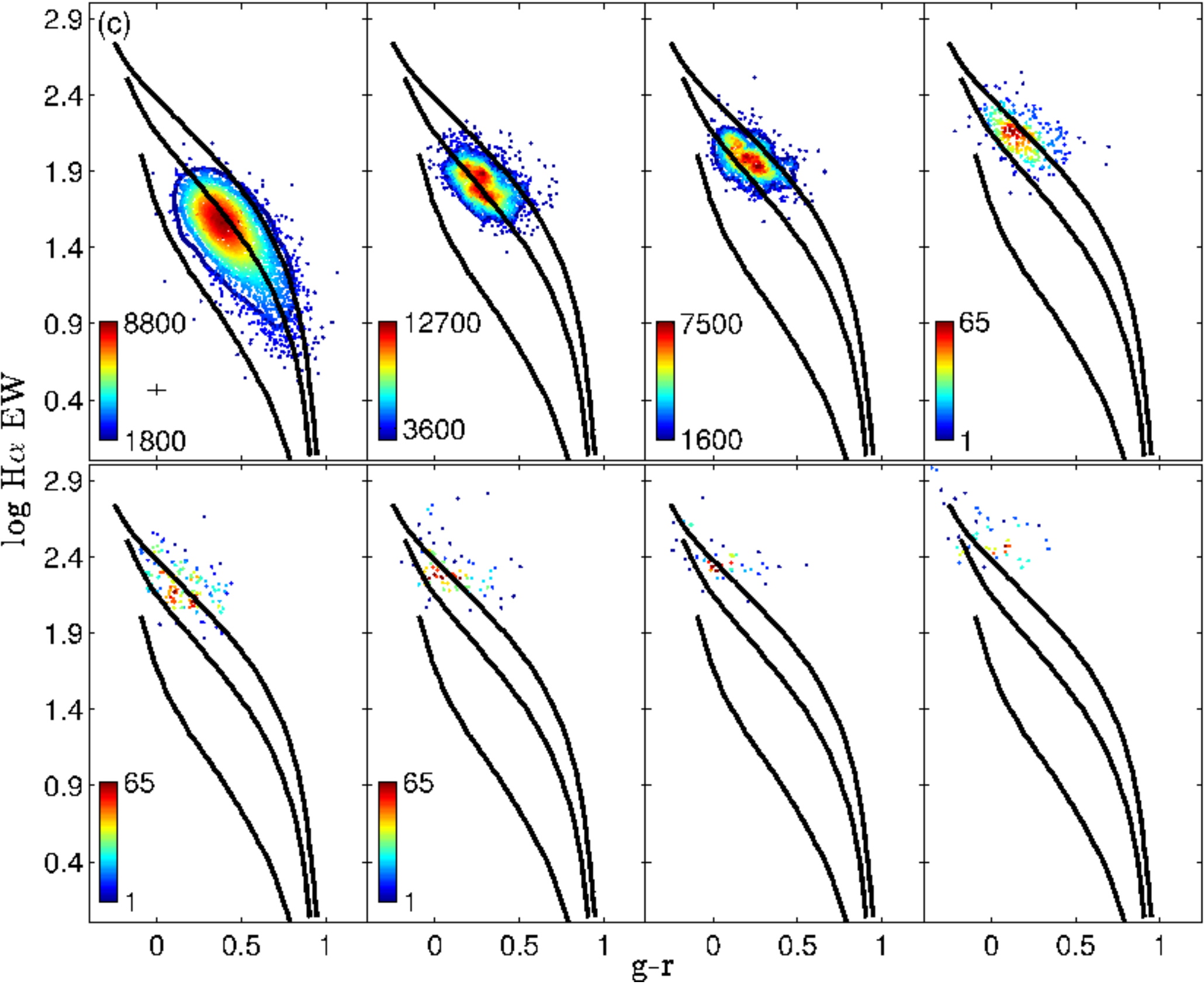}
\caption{Three independent volume limited samples $\langle M_r, z \rangle = -21, 0.29$ (a), $\langle M_r, z \rangle = -20.5, 0.19$ (b), and  $\langle M_r, z \rangle = -19.5, 0.13$ (c) divided into eight SFR sub--samples. A representative uncertainty on individual measurements is indicated by the error bars in each top--left panel. The IMF--SFR relationship is evident in each independent volume limited sample. The coloured contours are based on the data densities and the ranges of these densities are indicated alongside each colour bar.}
\label{fig:sfrBin_highestVolBin}
\end{center}
\end{figure*}

A  clear variation in IMF slope with the SFR of the host galaxy is evident in Figure~\ref{fig:sfrBins}, where the high star formation rate systems are characterised by a flatter IMF. 
In order to quantify this effect, the same analysis is performed using three independent volume limited samples, with absolute $r$--band magnitude ranges centered on M$_r=-21, 20.5$ and $-19.5$ shown in Figure~\ref{fig:sfr_z_mass}(b), where each sample is complete to a given $r$-band luminosity. This avoids the bias against lower SFR systems at higher redshifts, imposed by our optical/near--infrared magnitude and H$\alpha$\, flux--limited selection. Each sample is further divided into eight sub--samples based on SFR (Figure~\ref{fig:sfrBin_highestVolBin}). The clear progression towards a top--heavy, or flatter, IMF with increasing SFR is evident in all three independent volume limited samples.

The general trend measured here is that low SFR galaxies populate the lower right of the H$\alpha$ EW and colour plane, being characterised by Salpeter, or steeper,  IMF slopes. With increasing SFR, the galaxy population moves upwards to the left from a steep to a flat IMF track, implying a SFR dependence of the high--mass IMF slope. Given the low redshift range ($0<z\leq0.35$) of the sample, the observed IMF--SFR effect cannot be a result of merging systems since the merger rate is very low, $\sim 2\%$ \citep{PCL07, Lotz08} at these low redshifts, and the high--SFR systems are not dominated by mergers. 

\subsection{The effect of dust extinction}

\begin{figure*}
\begin{center}
\includegraphics[scale=0.44]{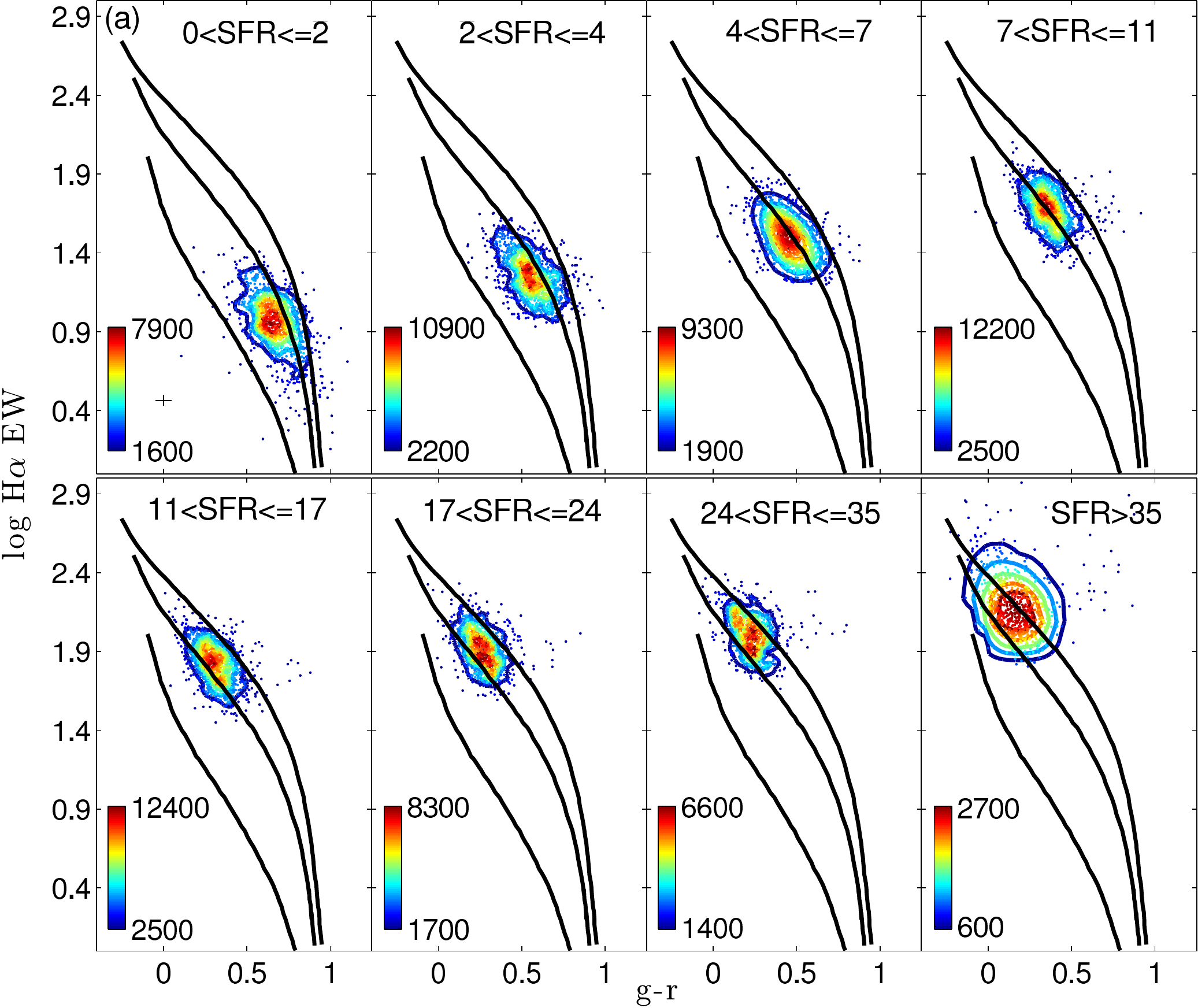}
\includegraphics[scale=0.44]{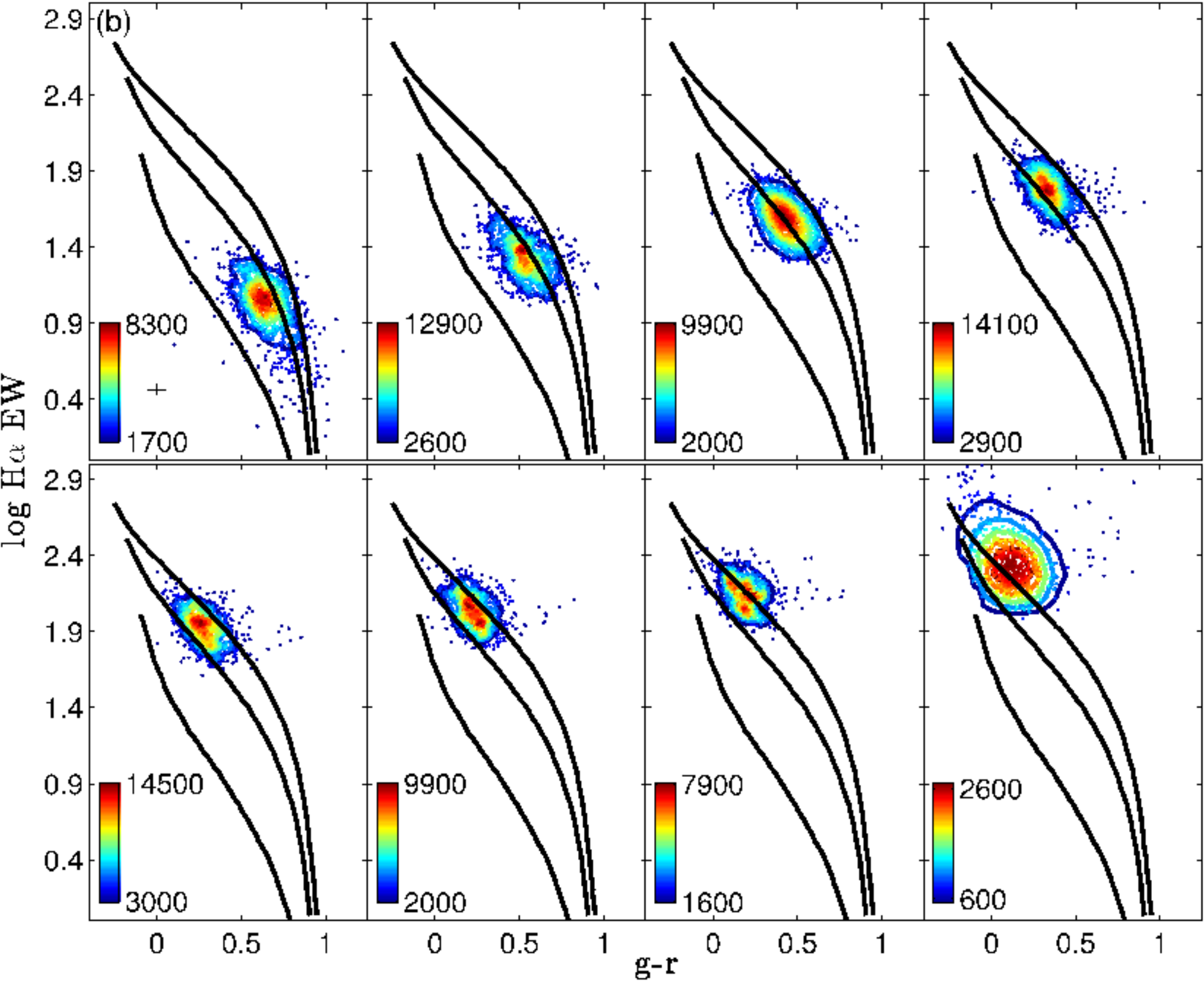}
\includegraphics[scale=0.44]{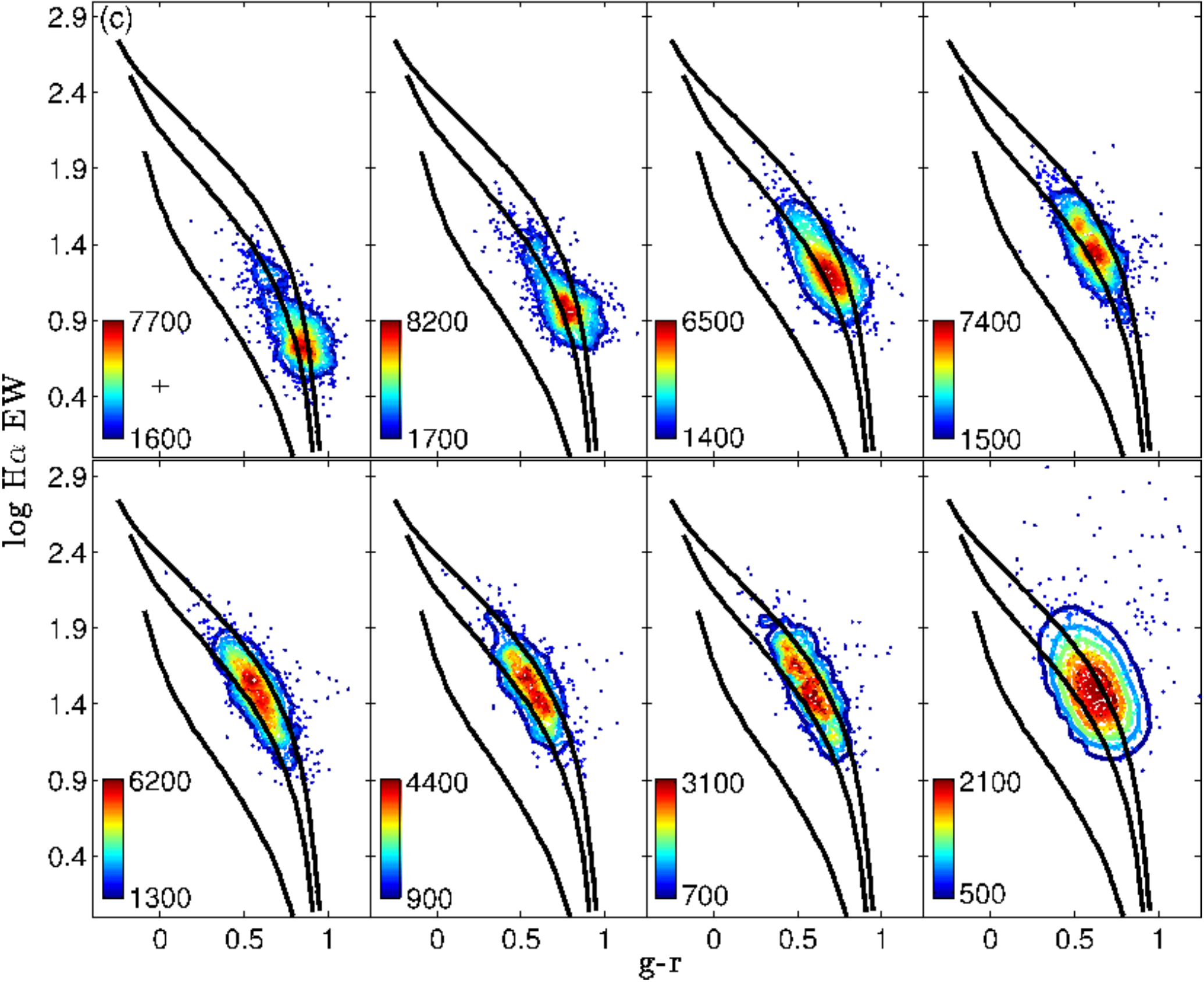}
\caption{$\langle M_r, z \rangle = -21, 0.29$ sample divided into eight SFR sub--samples. (a) Data corrected for obscuration using  \citet{Cardelli89} for emission line corrections and \citet{Calzetti01} for continuum corrections, (b) data corrected for obscuration using  \citet{FD05} as prescribed by \citet{WAS10} and  (c) data without any obscuration correction applied. Model tracks as in previous figures.}
\label{fig:dust_cor}
\end{center}
\end{figure*}

Figure\,\ref{fig:dust_cor} compares the uncorrected results with dust obscuration corrected using the \cite{Cardelli89} and \cite{Calzetti01} obscuration curve combination and the \cite{FD05} obscuration curve as given by \cite{WAS10}.

The effect of dust from each of the prescriptions investigated is to move data points parallel to the model evolutionary tracks. The progression of data orthogonal to the tracks, as SFR varies, is unlikely to be a consequence of erroneous dust obscuration corrections. Finally, increasing $\tau_b$ does not affect H$\alpha$ EW except at high inclinations, where H$\alpha$ is attenuated more than the $r$--band due to the low scale height. Therefore, if there is an increase in disk opacity ($\tau_b$) as a function of SFR, this effect would not cause a systematic shift of data points orthogonal to the model tracks, mimicking the trend with SFR presented in this paper. Any such effect due to the presence of dust causes the raw data points to move downwards parallel to the model tracks.

\subsection{Addressing the systematics}

Here we explore how the modification of the other free parameters of the population synthesis models, as well as the introduction of additional models, affects our results. 

\subsubsection{Effects of modifying the free parameters}
\begin{figure*}
\begin{center}
\includegraphics[scale=0.5]{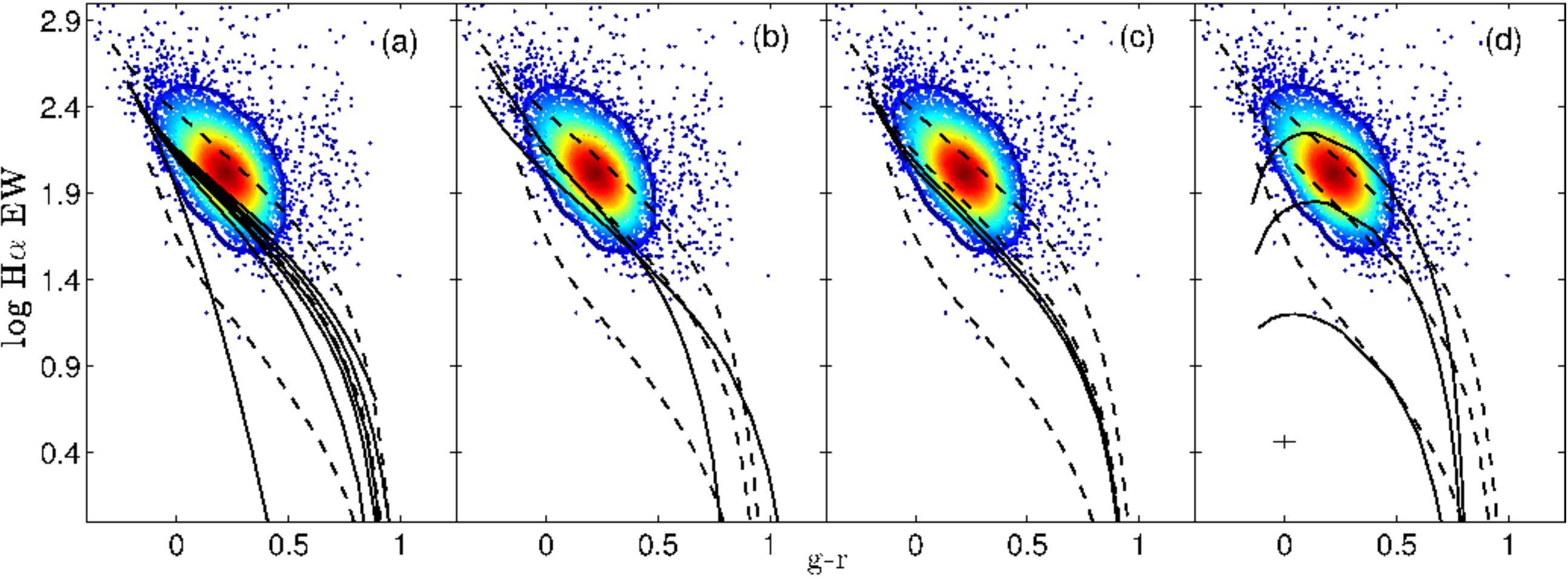}
\caption{(a) The effect of varying $\tau$ values on the model evolutionary track $\alpha = -2.35$ with $\tau =1.1$ Gyr.  $\tau = 0.1, 0.5, 0.8, 1, 1.5, 2.5, 13$ Gyr $e$-folding times are shown as solid lines. The $e$-folding time increases from left to right and all solid tracks have input IMF slope of $\alpha=-2.35$. (b) The effect of metallicity on the model tracks. The two tracks with solid lines have $\alpha=-2.35$, the same as the middle dashed lined track, which has $Z=0.02$, i.e. solar metallicity. The solid line extending above the dashed line at high $g-r$ and low H$\alpha$ EW has $Z=0.05$ and the other has $Z=0.005$. (c) The variation in the universal IMF track, i.e. $\alpha=-2.35$ ($0.5<M/M_{\odot}<120$), with respect to two different assumed stellar upper mass limits. The stellar upper mass limits of the three tracks are 120 (dashed track), 100, 90 $M_{\odot}$ (solid lines), from top to bottom. (d) Solid lines denote Maraston model tracks. The dashed tracks in all panels are the three main tracks shown in previous figures. The time spans of all of the tracks are 100 Myr to approximately 13 Gyr, running from top left to bottom right. Also shown in each panel is the positions of the high SFR ($>13$\,M$_{\odot}\,yr^{-1}$) galaxies in the GAMA sample.} 
\label{fig:IMF_effects}
\end{center}
\end{figure*}

The three model evolutionary tracks plotted in previous figures and shown as dashed lines in Figure\,\ref{fig:IMF_effects} are generated assuming an exponentially decreasing star formation history with $e$-folding time of $1.1$\,Gyr, $Z=0.02$ (solar metallicity) and an upper stellar mass limit of $120$\,M$_{\odot}$ for different input IMFs. Figure\,\ref{fig:IMF_effects} explores the effects of varying these free parameters in generating model evolutionary tracks and the use of other models. 

The modification of the free parameters of P\'EGASE model does change the position of the evolutionary tracks. These changes however, largely move the tracks towards the bottom of the plots. Although this can potentially explain data in lower--left of the diagram (i.e.\,low SFR galaxies) without resorting to evolving IMFs, in none of the cases can it explain the data in the upper right of the diagram, as shown in Figure\,\ref{fig:IMF_effects}(a--c). In other words an evolutionary track with an input flat IMF is required to describe the highest SFR sources, while the positions of low SFR galaxies can be explained either by using evolutionary tracks with input Salpeter--like or steep IMFs and $\tau=1.1$\,Gyr or with a top--heavy IMF with varying $e$-folding times. 

\subsubsection{Different population synthesis models}

Figure\,\ref{fig:IMF_effects}(d) shows the positions of the \cite{M05} model evolutionary tracks and the highest SFR objects in the sample with respect to the P\'EGASE tracks. The Maraston tracks are generated based on the same input criteria used to generate the three P\'EGASE tracks. In this part of the analysis we investigate the effect of the inclusion of  the Thermally Pulsating--Asymptotic Giant Branch (TP--AGB) phase on the model evolutionary tracks. Note that the TP--AGB phase is included in the P\'EGASE models; however, it may be the case that P\'EGASE models may not be adequate at describing real stellar populations with TP--AGB stars \citep{M05}.

The current version of the \cite{M05} models only provides continuum fluxes and thus colours for a given IMF, metallicity and star formation history. The H$\alpha$ EWs are calculated by combining the H$\alpha$ line flux from P\'EGASE with the continuum flux from the Maraston models at each common time step. The presence of TP--AGB stars has a significant effect on the colour of a galaxy; however their contribution to H$\alpha$ emission is expected to be negligible. The turn--over of the Maraston tracks evident in Figure\,\ref{fig:IMF_effects} occurs at around the age ($0.2\leq t/Gyr\leq2$), when the TP--AGB contributions become important. The difference in model tracks at this age arises due to the higher continuum fluxes given by the Maraston model in comparison to the P\'EGASE model. Despite this difference at early times, it is clear that, even using the Maraston models, the higher SFR systems favour flatter IMF slopes. 

 We further tested the STARBURST99 models \citep{starburst99} to explore whether there is a variation in H$\alpha$ emission predicted by different models. We found that the use of STARBURST99 or P\'EGASE gave essentially identical results for H$\alpha$ emission. We conclude that the choice of population synthesis models does not significantly alter our main results.

\subsection{SDSS vs GAMA}

\begin{figure*}
\begin{center}
\includegraphics[scale=0.44]{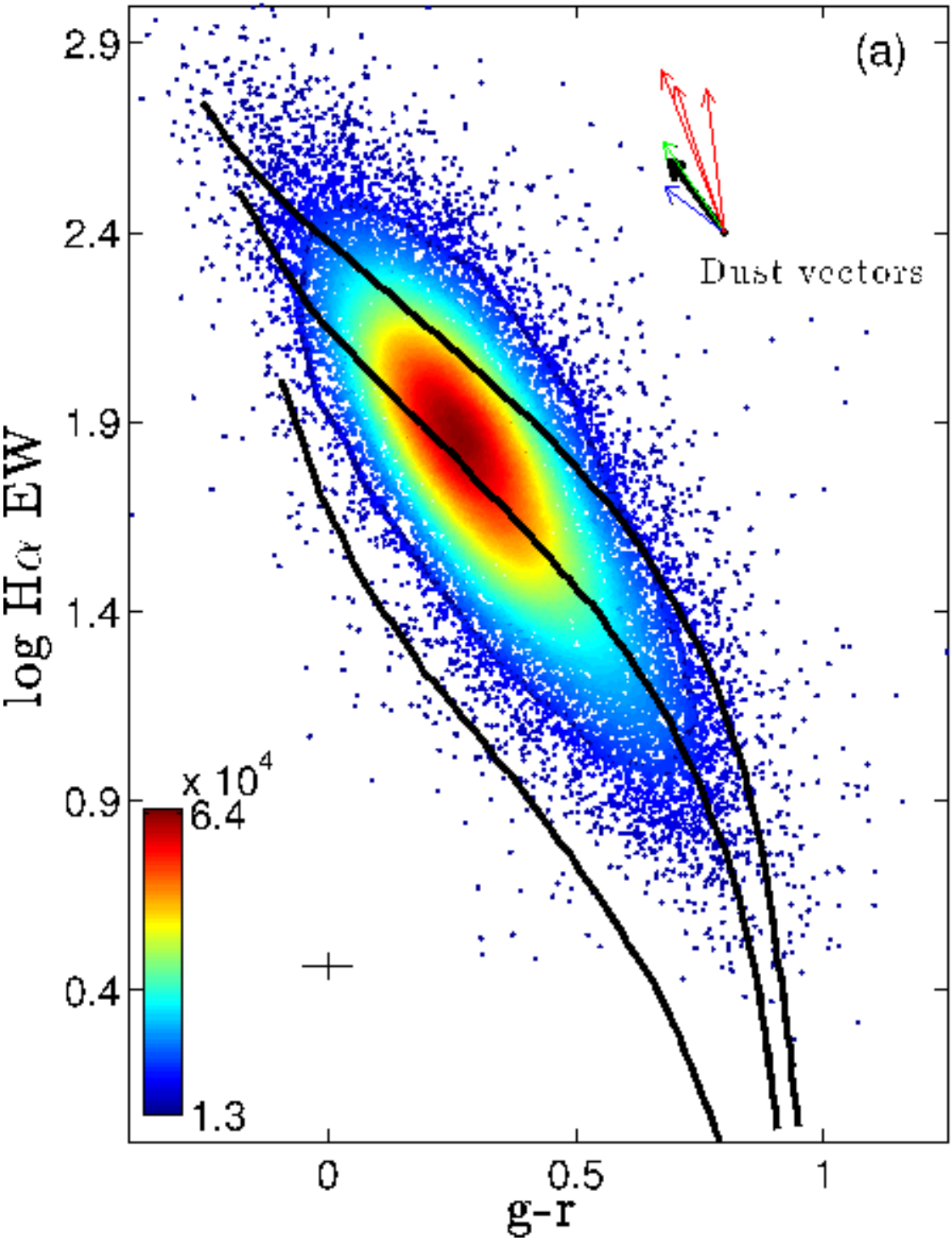}\\
\includegraphics[scale=0.45]{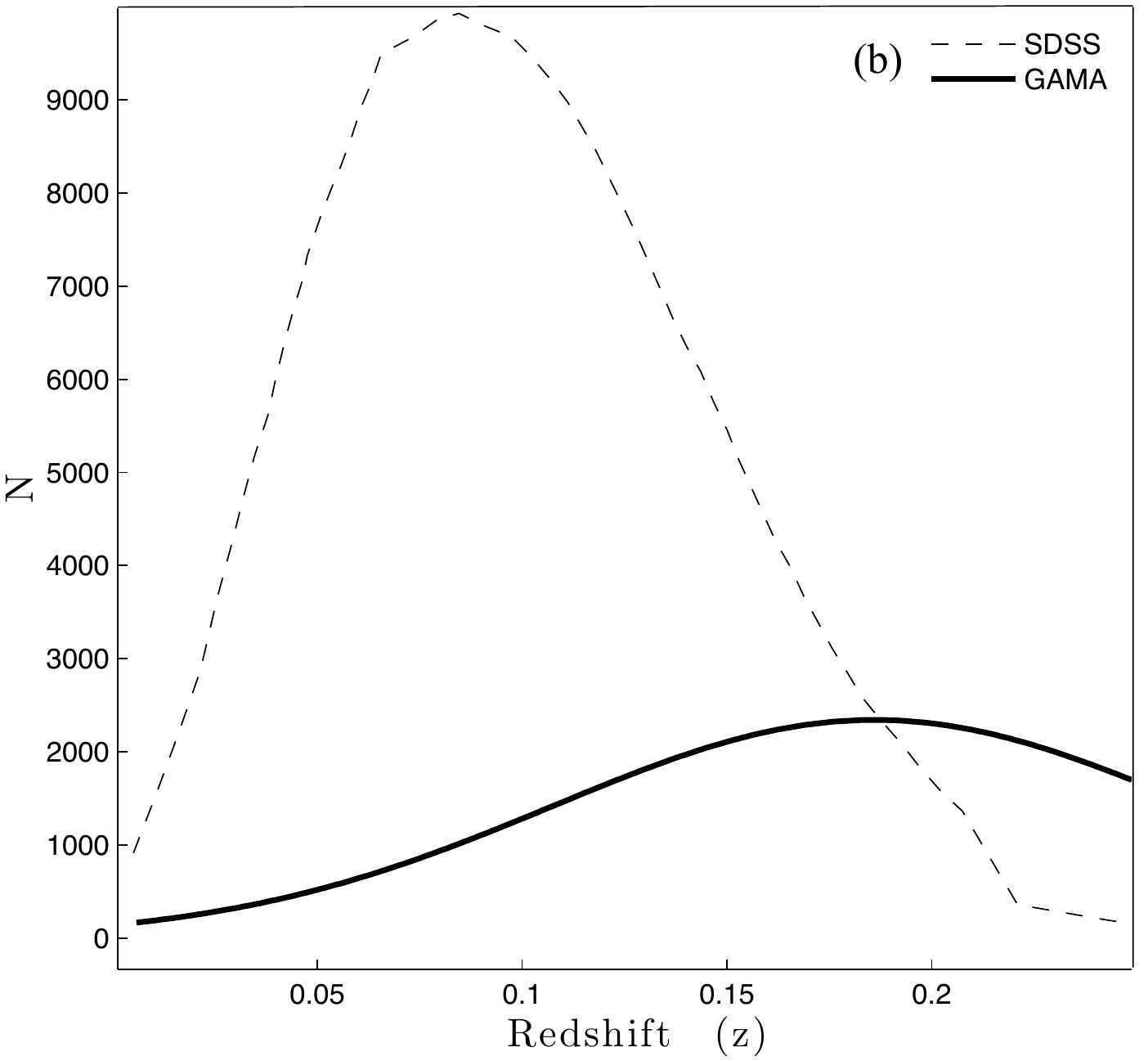}
\includegraphics[scale=0.4]{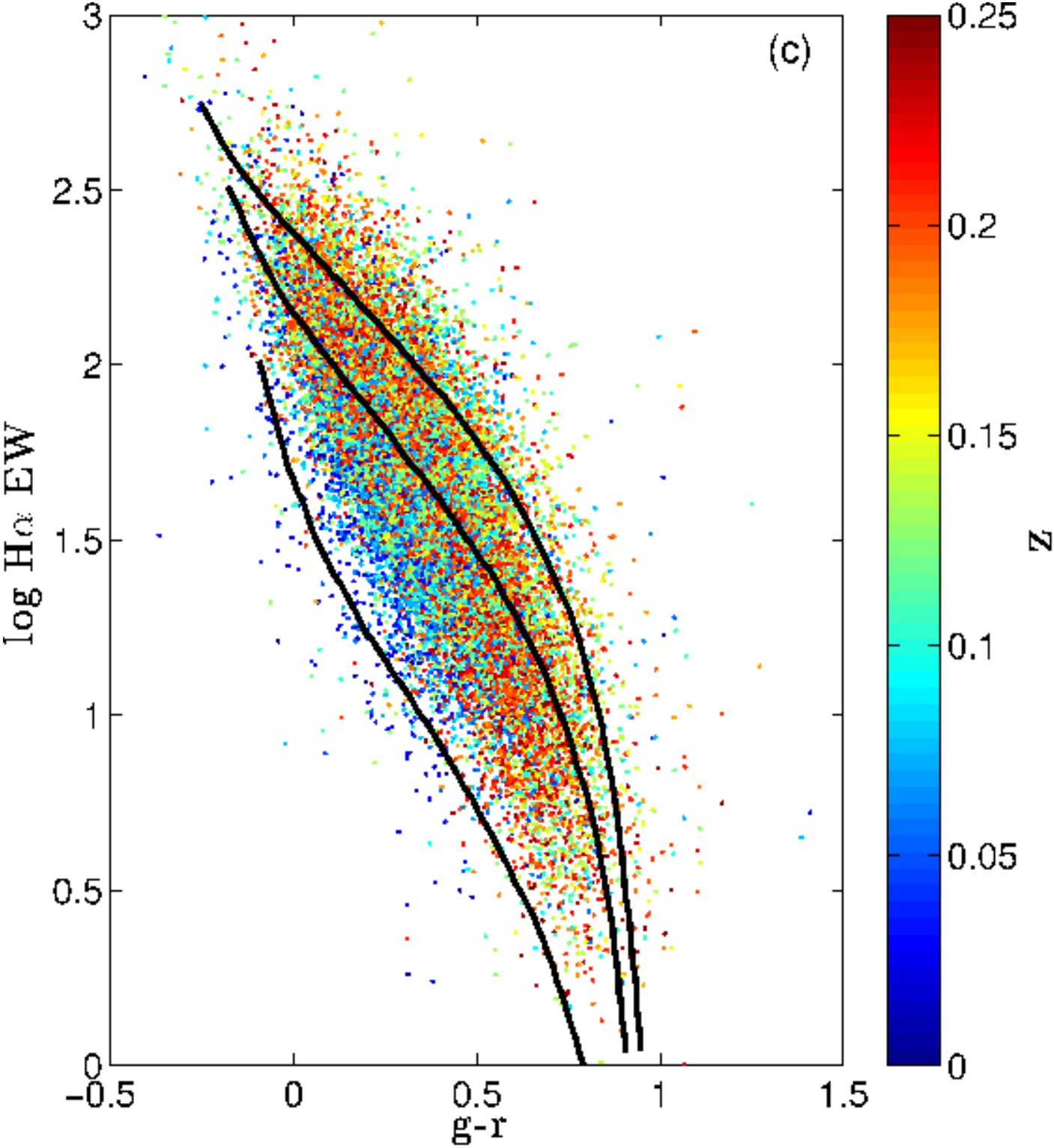}
\caption{(a) The distribution of the GAMA galaxy sample spanning $0<z\leq0.25$. The model tracks from the top to bottom have the following IMF slopes, $\alpha=-2, -2.35, -3$. This figure is similar to Figure 1 of \citet{HG08}; the scale and model IMF tracks of both figures are the same. (b) The  distribution of the redshift of the SDSS main sample and the GAMA sample for $0<z\leq0.25$. (c) The GAMA sample colour coded according to $z$. This sample covers the range $0<z\leq0.25$. It is clear that the reason the peak in the data density in Top--panel lies higher than in the corresponding diagram from \citet{HG08} is because of the higher redshift range sampled.}
\label{fig:zCut}
\end{center}
\end{figure*}

A recent study by \citet{HG08}, using a sample of $\sim130\,603$ SDSS galaxies spanning $0<z\leq0.25$, found that the majority of the galaxies in their sample, which is dominated by low--luminosity systems, prefer steep IMFs. The same redshift limits are used with the GAMA sample in order to compare the GAMA and SDSS galaxy distributions. The GAMA galaxy distribution is shown in Figure\,\ref{fig:zCut}(a), and the majority of the sample indicates a preference for a flatter IMF. This is a consequence of the different redshift distributions of the two galaxy surveys, as demonstrated in Figure\,\ref{fig:zCut}(b,c).  GAMA has $\langle z \rangle \approx 0.2$ within this range, while SDSS has $\langle z \rangle \approx0.08$. The GAMA sample is dominated by relatively high $z$ galaxies, with higher SFRs, and which we have demonstrated are those that favour a flatter IMF slope.  

\section{star formation bursts}

A sudden burst of star formation on top of an otherwise exponentially declining star formation history would give rise to a large EW (i.e.\,increased SFR) and make the galaxy appear blue for a short period of time. The effect of a burst is therefore to push the EWs of galaxies with a Salpeter IMF to high H$\alpha$\, EW and low $g-r$, potentially leading to the erroneous inference of a flatter IMF for such a galaxy if only its position in the H$\alpha$\, EW and $g-r$ plane is considered. This was explored in some detail by \citet{HG08}, who argue bursts are unlikely to explain the variation in IMF, as the necessary bursts would have to be unrealistically coordinated in time to produce the observed galaxy colours. Here we describe several methods that we employed to rule out bursts as the possible source of the observed IMF--SFR dependency. 

\subsection{Mass--doubling times}
\begin{figure*}
\begin{center}
\includegraphics[scale=0.4]{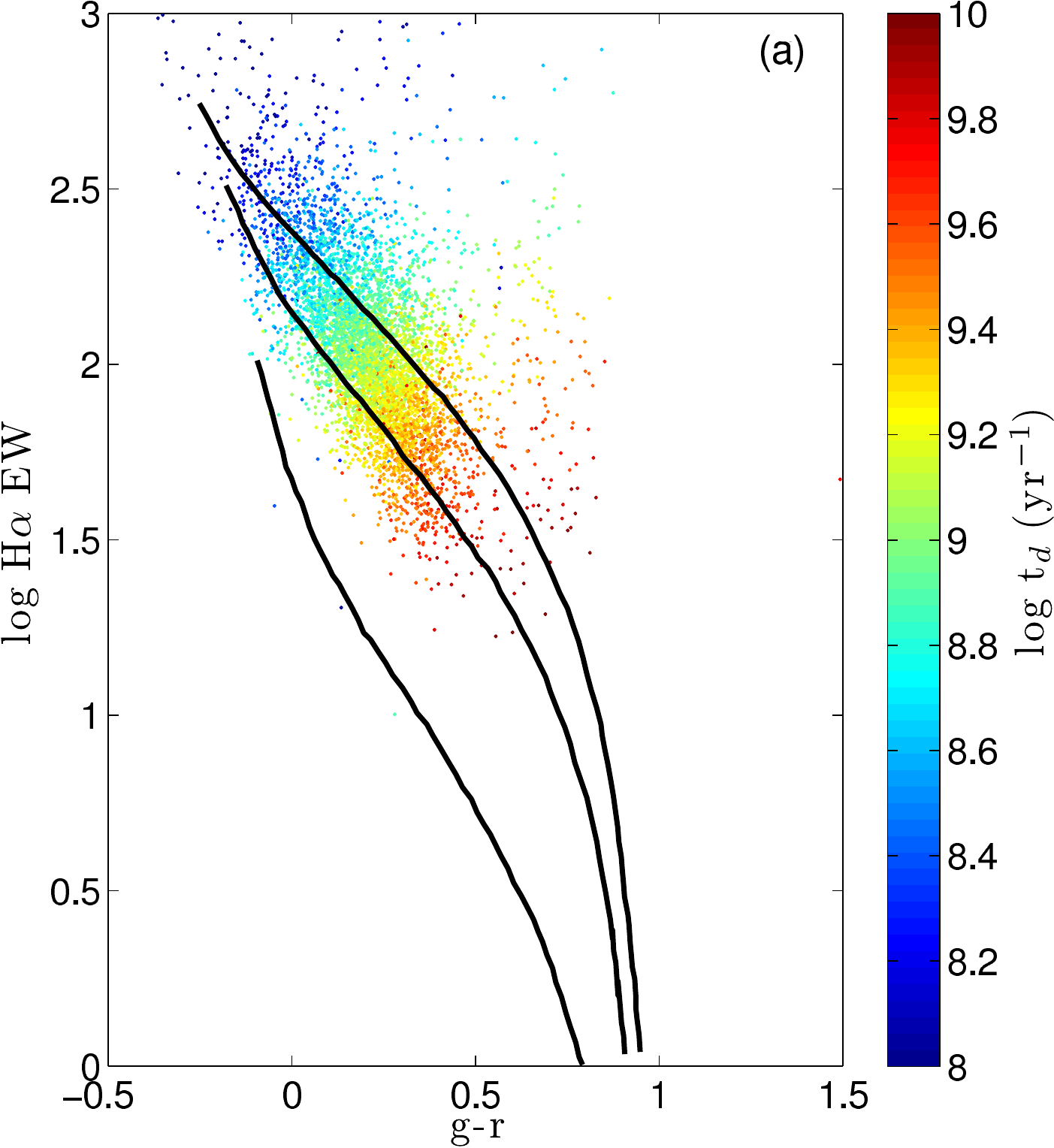}
\includegraphics[scale=0.42]{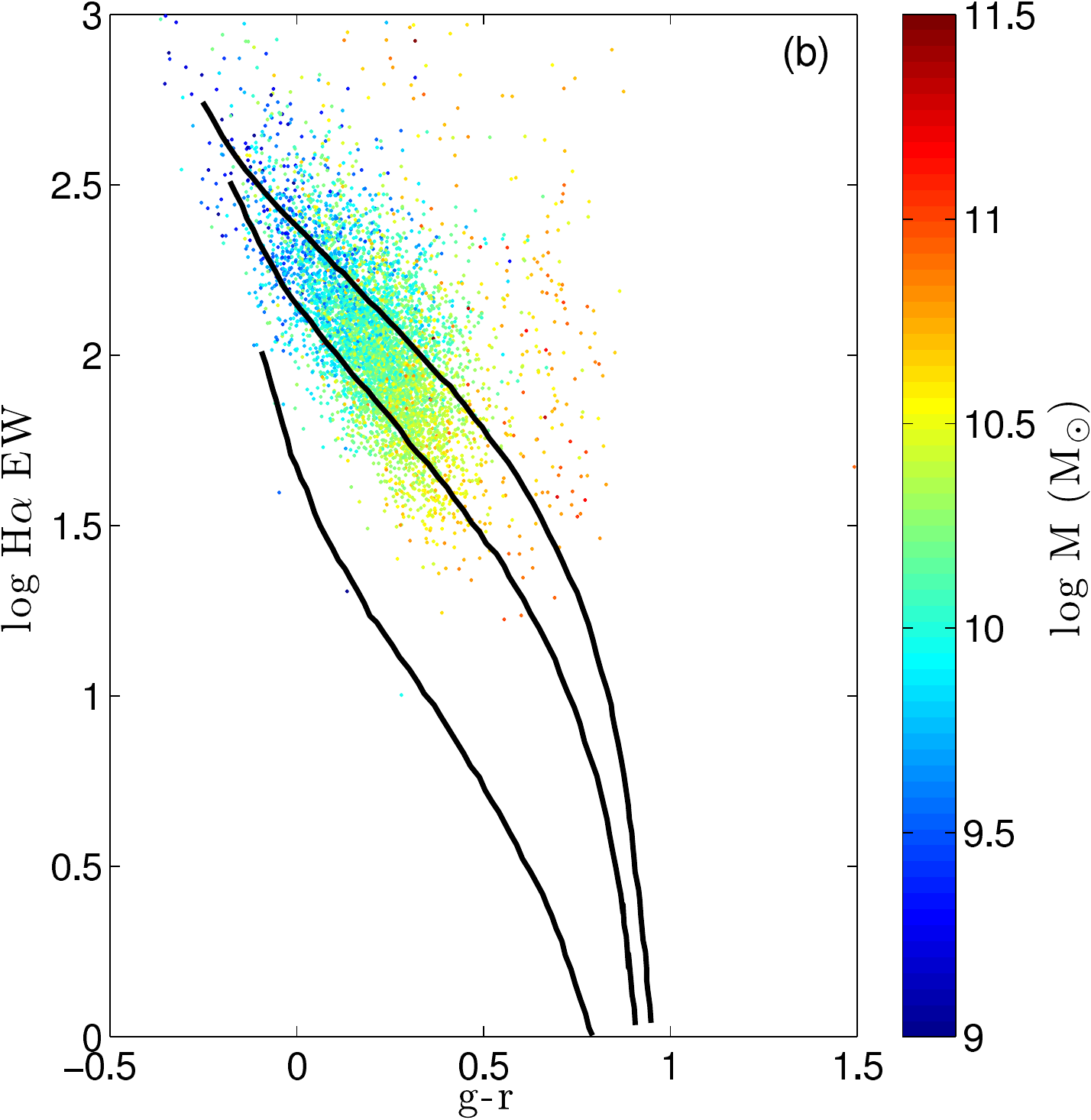}
\caption{(a) The H$\alpha$\, EW and $g-r$ for the highest SFR sub--sample of galaxies (SFR (M$_{\odot}$yr$^{-1}$)$\geq 13$) colour coded according to their mass--doubling times, (b) and according to their masses.}
\label{fig:t_d}
\end{center}
\end{figure*}
The mass--doubling times (t$_d$) of galaxies provide a method of isolating those galaxies undergoing a star--bursting phase. The mass--doubling time, used to calculate the t$_d$ values for the highest SFR sub--group of galaxies, is defined as  $t_d = M_*/(0.5\times SFR)$ \citep{Noeske07}, and provides a time scale within which the current SFR would produce the observed stellar mass (M$_*$).  Folded within the constant term ($1-R=0.5$) is the IMF dependent gas recycling factor, which determines the fraction of gas recycled into the interstellar medium. Galaxies with t$_d$ significantly shorter (t$_d<0.1-1$\,Gyr) than the adopted ages of galaxies are potential bursts. A galaxy experiencing a burst of star formation has a relatively high specific SFR (i.e.\,SFR per stellar mass) and hence corresponds to a low t$_d$.  The derived t$_d$ values for the highest SFR sub--group presented in Figure~\ref{fig:t_d}(a) show that the majority of high--SFR galaxies have relatively large t$_d$ values (t$_d>1$ Gyr), indicating that they are not currently in a starburst mode. In addition, both t$_d$ and mass vary smoothly {\it along} the P\'EGASE model tracks from top--left to bottom--right, without the vertical gradients in colour that would be expected from starbursts. The IMF dependence in the estimation of t$_d$ will be small. Since $t_d \propto M/SFR$, the dependencies of mass and SFR on the IMF  largely cancel (although not entirely, as these dependencies are not identical). The dependency of $R$ on the IMF is likely to have a small effect on the t$_d$ values, as $R$ is inversely related to the slope of the IMF \citep[e.g.\,$R\approx0.28$ for $\alpha\approx-2.35$ and $R\approx0.56$ for $\alpha\approx-2.15$;][]{HB06}, such that t$_d$ increases if a flatter IMF is assumed and vice versa.
\begin{figure*}
\begin{center}
\includegraphics[scale=0.43]{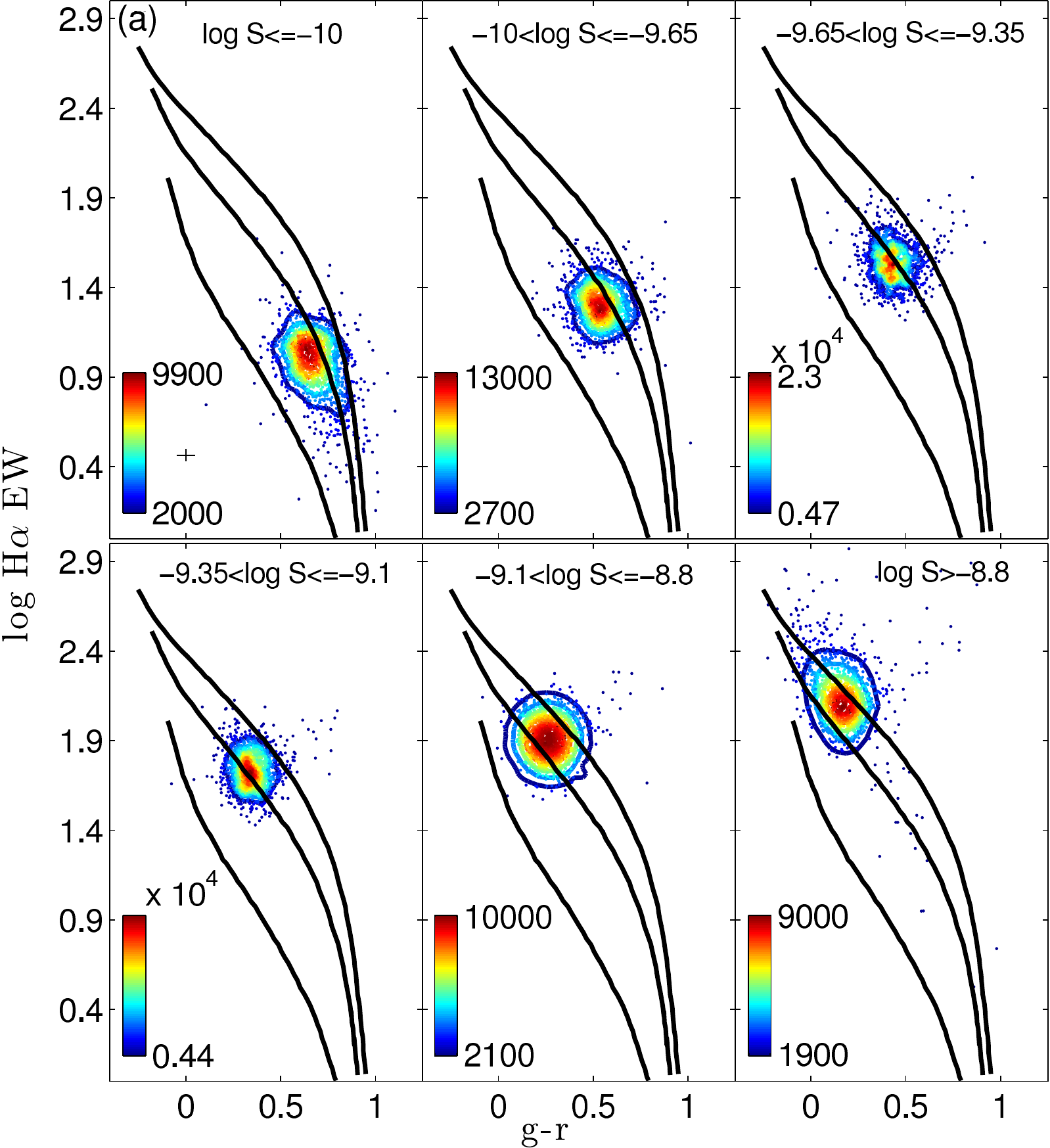}
\includegraphics[scale=0.47]{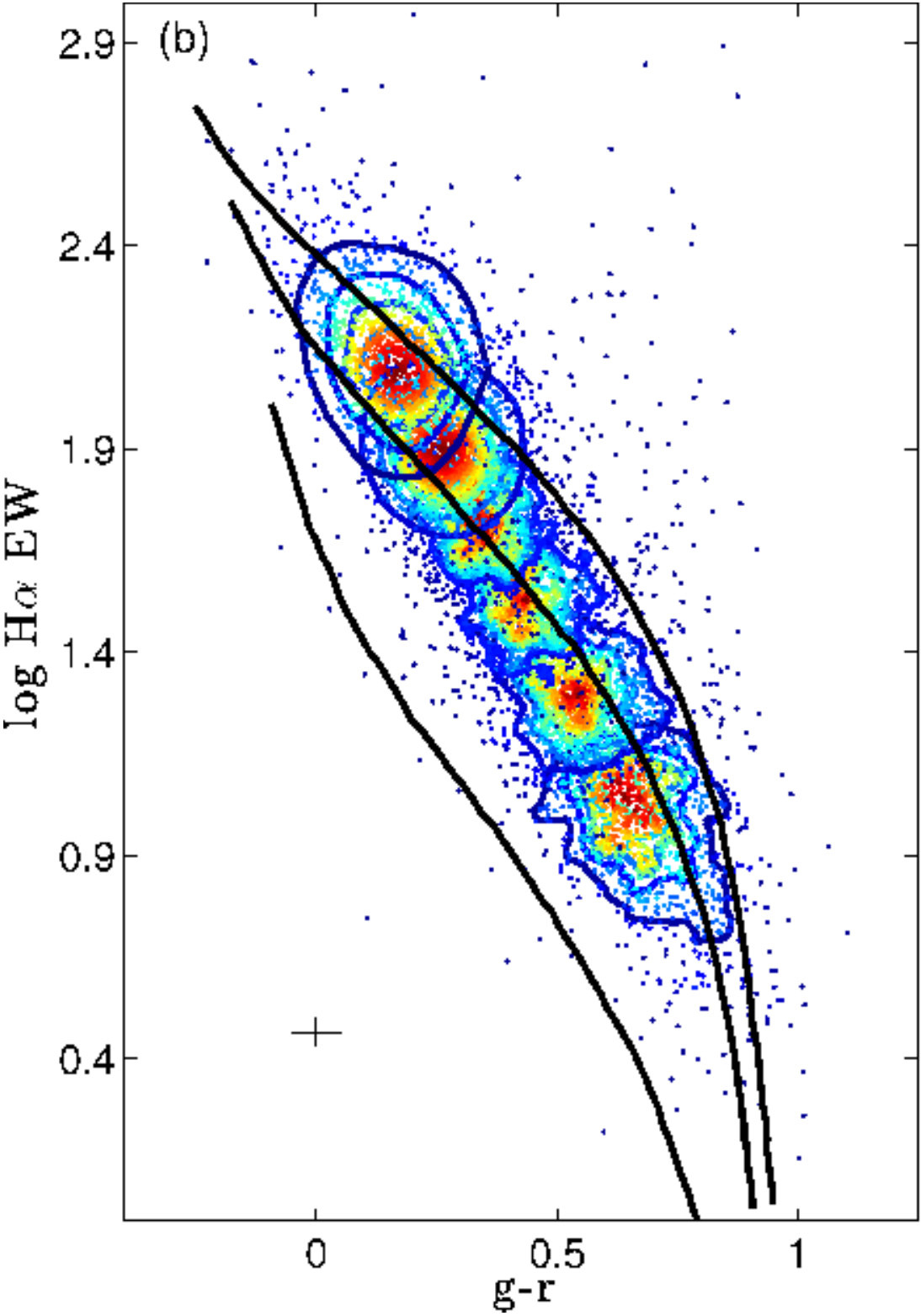}
\caption{ (a) The highest volume limited sample $M_r=-21$ divided into sub--groups of `specific' SFR (S denotes the specific SFR in units of yr$^{-1}$). Specific SFR increases from left--to--right, top--to--bottom, and (b) summaries this trend clearly.}
\label{fig:SSFR}
\end{center}
\end{figure*} 

According to the staged galaxy formation scenario of \citet{Noeske07}, the high specific SFRs (low t$_d$) of most low mass galaxies (M$\leq10^{10}\,M_{\odot}$) are not indicative of evolved galaxies experiencing a starburst.  In fact an initial burst followed by gradual decline seems to be the favoured mode of star formation. However, this initial dominant burst of star formation is pushed towards later redshifts for less massive galaxies. This further supports our argument that the highest star formation rate objects of our sample are not starbursts but quiescently evolving galaxies inherently preferring a flatter IMF in comparison to low SFR systems. We reiterate that the smooth IMF--SFR variation is evident in all three independent volume limited samples.  The observed trend is thus strong evidence for an IMF--SFR relation. 

 We next measure the dependencies on specific star formation rate and star formation rate surface density, to allow an exploration of the most likely underlying dependency of the IMF.

\subsection{Specific star formation rates}\label{sec:SSFR}

Specific SFR is calculated as the SFR per unit stellar mass. Figure\,\ref{fig:SSFR} shows the variation with specific SFR for the highest redshift volume--limited sample. This demonstrates a very similar result to that found above for absolute SFR, in the sense that systems with higher specific SFR also prefer flatter IMF slopes. Note that even the highest specific SFR systems here still have mass doubling times of the order 1\,Gyr, a consequence of their high average masses (see also Figure\,\ref{fig:t_d}). The smooth decline along the model evolutionary tracks with decreasing specific SFR, evident in Figure\,\ref{fig:SSFR}, is consistent with smooth star formation histories. Similar results are seen in the other two volume limited samples.

\subsection{Star formation rate surface density}\label{sec:SFRD}

\begin{figure}
\begin{center}
\includegraphics[scale=0.37]{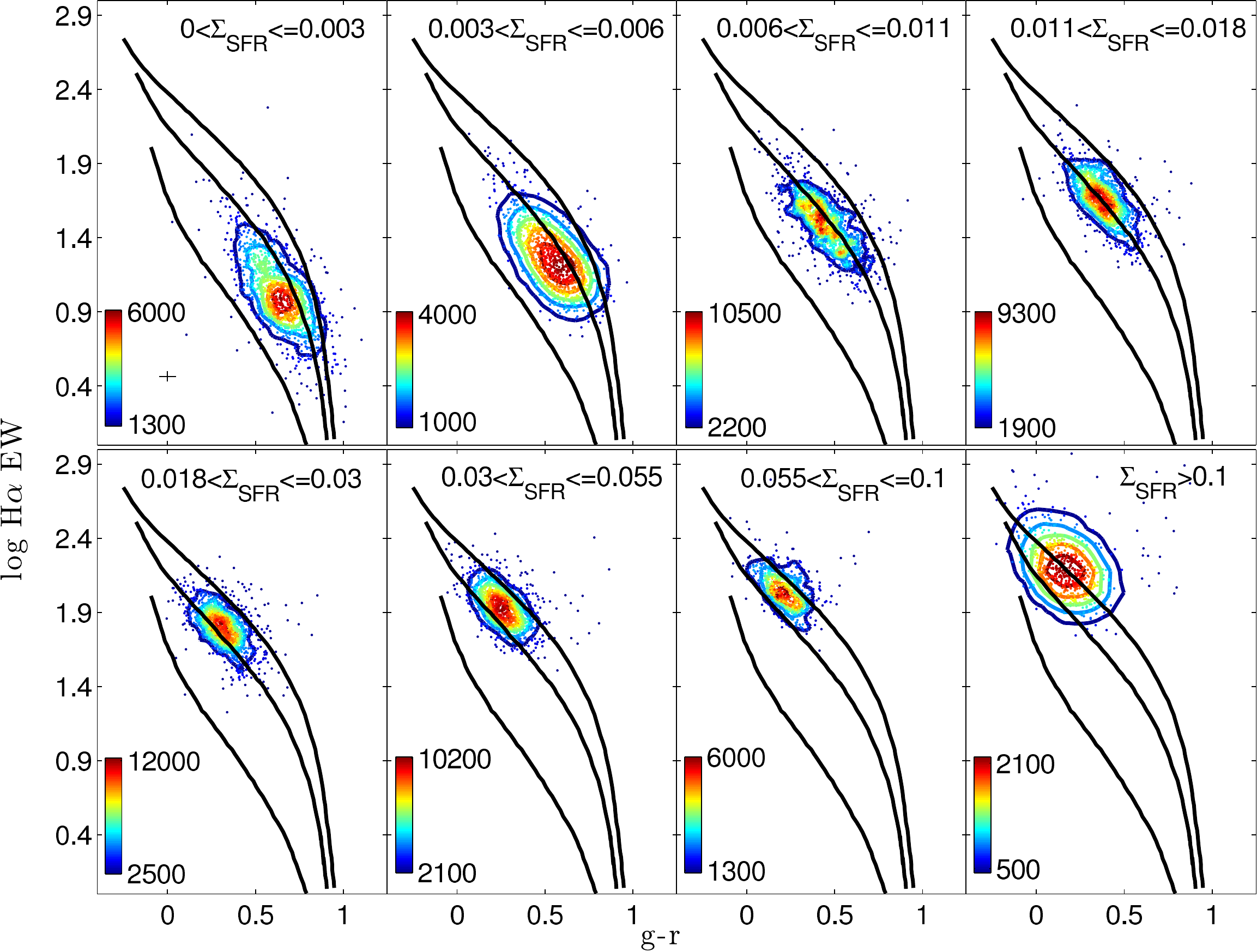}
\caption{The highest volume limited sample $M_r=-21$ divided into sub--groups of star formation rate surface density (i.e.\,SFR per unit area, in units of M$_{\odot}$yr$^{-1}$kpc$^{-2}$). The SFR surface density increases from left to right, top to bottom.}
\label{fig:SFRD}
\end{center}
\end{figure}

Figure~\ref{fig:SFRD} explores the trends with respect to SFR per unit surface area in M$_{\odot}$yr$^{-1}$kpc$^{-2}$. The Petrosian radius, which is derived from the surface brightness profile of the galaxy, provides a measure of the angular size of the galaxy. Assuming the Petrosian radius to be the radius of a circle, the surface area projected on the sky can be calculated, from which the SFR per surface area can be determined. However, galaxies are not typically circular, and there is a range of morphological types, with ellipticals, spirals and irregulars all present in the local galaxy population. The surface areas projected on the sky by these types are best described using ellipses. The effect of assuming a circular form is to reduce the SFR per area. This results in moving objects between the SFR surface density bins shown in Figure~\ref{fig:SFRD}. Given the large ranges of SFR per surface area of the eight bins, the interchange of objects between bins is unlikely to be significant, and is dominated by objects within the scatter outside the lowest density contour. The central contours showing the highest density regions are not affected by a low interchange of objects. Hence, we can categorically say that all objects in the highest SFR per area bin have high SFR surface density and certainly prefer a shallow IMF slope. Furthermore, the SFR-IMF dependency shown is not primarily due to galaxy sizes, e.g., massive galaxies having higher SFR due to the SF processes being distributed over a larger area than the less massive systems. The other volume limited samples follow similar trends as shown in Figure\,\ref{fig:SFRD} for the highest redshift volume limited sample.

 These results demonstrate a qualitatively similar dependence of the inferred IMF slope on SFR, SSFR and SFR surface density. The following section quantifies these dependencies, allowing us to identify which is likely to be the more fundamental.

\section{The fundamental IMF dependency}

Here we attempt to understand the underlying dependencies that define the shape of the IMF for galaxies and identify which is the more fundamental driver of IMF variations.  We have explored the variations with respect to SFRs, specific SFRs and SFR surface density. These cases are described in \S\ref{sec:SFR}, \S\ref{sec:SSFR} and \S\ref{sec:SFRD}.

\begin{figure*}
\begin{center}
\includegraphics[scale=0.37]{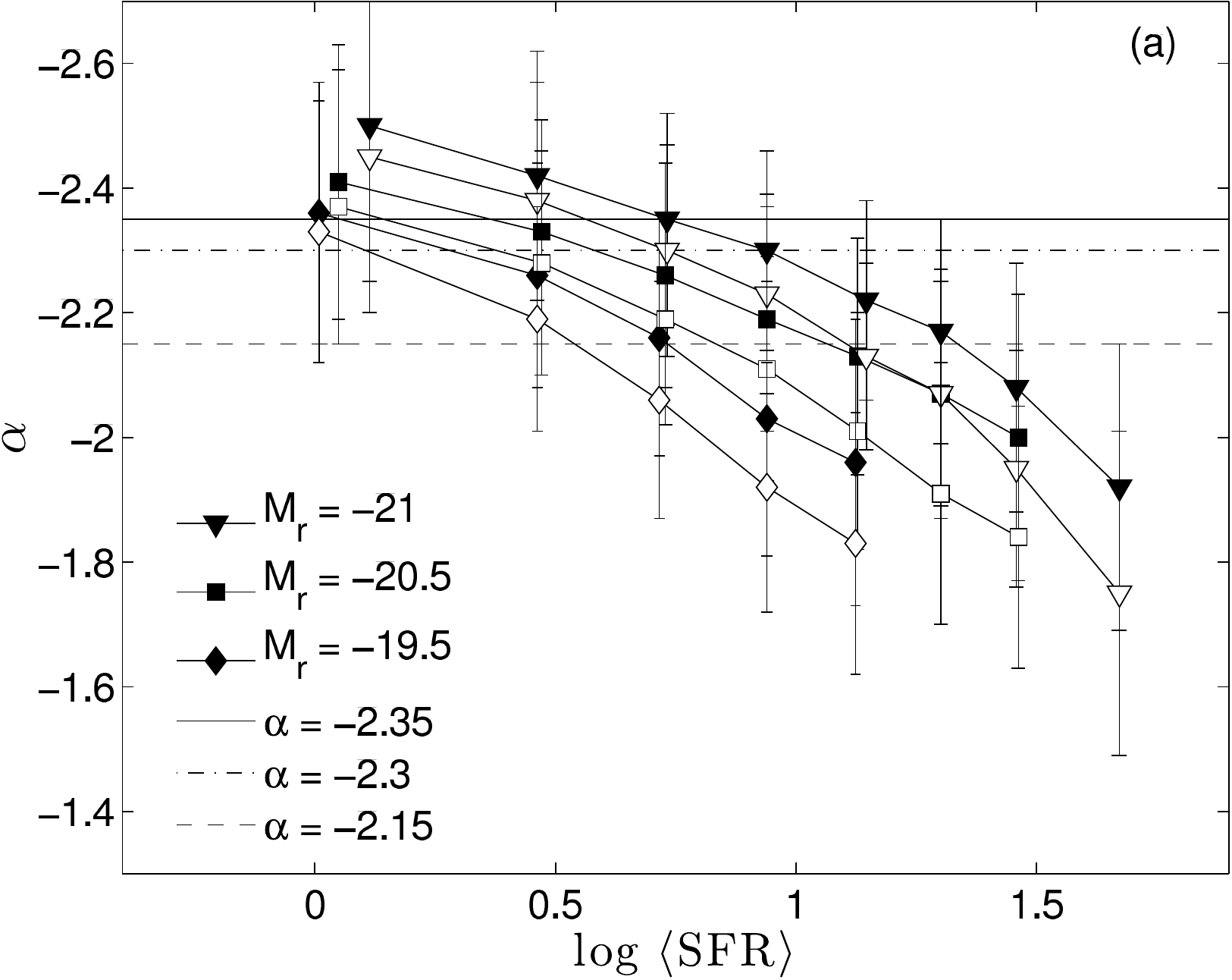}
\includegraphics[scale=0.37]{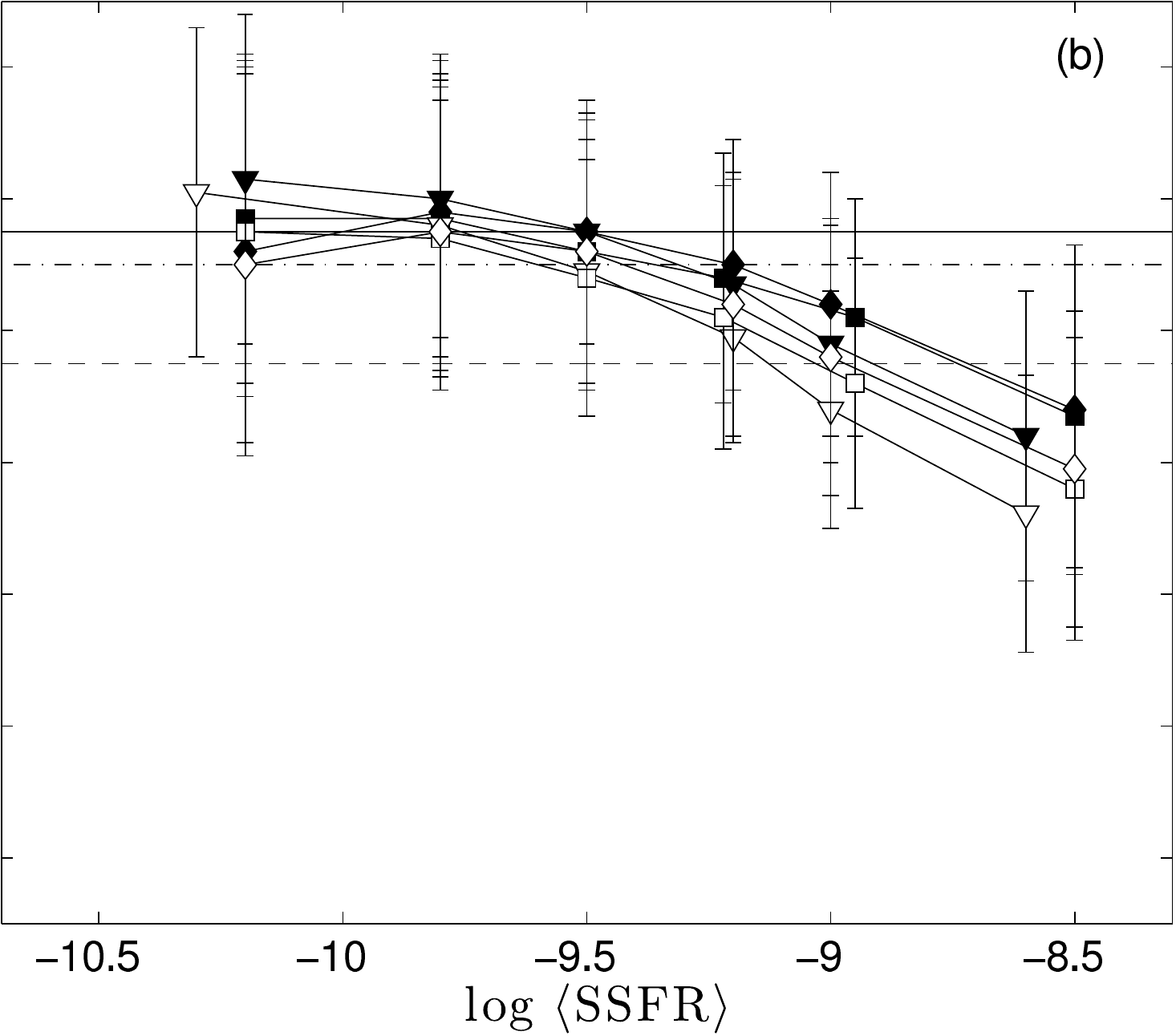}
\includegraphics[scale=0.37]{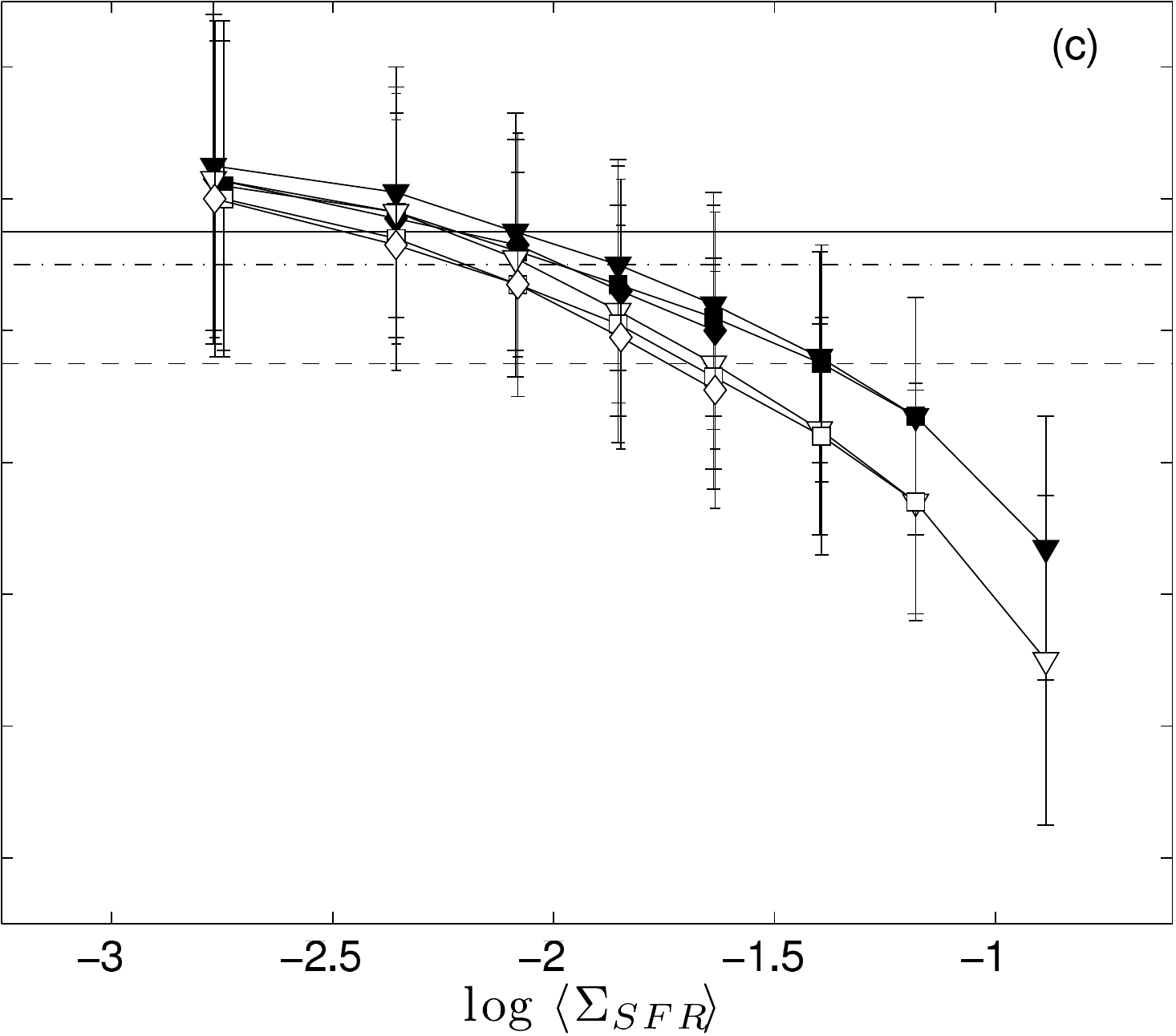}
\caption{{The best fit IMF slope for each of the (a) SFR (Figure~\ref{fig:sfrBin_highestVolBin}), (b) specific SFR (Figure~\ref{fig:SSFR})  and (c) SFR surface density (Figure~\ref{fig:SFRD}) sub--groups of the three volume limited samples used in this study. The filled symbols denote results when using the \citet{Calzetti01}/\citet{Cardelli89} dust corrections and open symbols represent the \citet{FD05} dust corrections. The solid horizontal line indicates a Salpeter slope, the dot--dashed line indicates a Kroupa (2001) high--mass slope of $\alpha=-2.3$ and the dashed line dictates the \citet{BG03} IMF slope. }}
\label{fig:IMF_slopes}
\end{center}
\end{figure*}

The best fit IMF slope ($\alpha$) is determined for each SFR, specific SFR and SFR surface density sub--group of the three volume limited samples. This is essentially the IMF of the P\'EGASE model track closest to the region of highest data density.
A library of model evolutionary tracks with input IMFs with high--mass slopes ranging from $\alpha = -1.45$ to $-3.05$ in increments of 0.1 is used to determine the best--fit IMF for each galaxy. The relationship between the IMF slope and mean property (i.e.\, SFR, specific SFR and SFR surface density) of objects in the respective sub--group is shown in Figure~\ref{fig:IMF_slopes}. The error in $\alpha$ is the standard deviation of the data in each respective sub--sample.
The filled symbols in Figure\,\ref{fig:IMF_slopes} correspond to the best fit $\alpha$ slopes if \citet{Calzetti01}/\citet{Cardelli89} dust obscuration curves are used to correct the data, while the open symbols show the effect if the \cite{FD05} dust obscuration curve is used instead. As shown in Figure\,\ref{fig:dust_cor}, dust obscuration corrections based on the \cite{FD05} curve seem to exaggerate the SFR--IMF trend. This can be seen explicitly in Figure 13.

The trend to steeper IMF slopes with brighter M$_r$ at a fixed SFR evident in Figure\,\ref{fig:IMF_slopes}(a), is likely to be a consequence of higher luminosity systems having higher masses. For a fixed SFR these systems have smaller specific SFRs, and we have also demonstrated (Figures~\ref{fig:SSFR} and \ref{fig:IMF_slopes}(b)) that galaxies with higher specific SFR have flatter IMF slopes. Figure\,\ref{fig:IMF_slopes}(b) highlights the result that, when accounting for the effect of galaxy mass, there is a surprisingly tight relation between inferred IMF slope and specific SFR. It is possible that there may also be metallicity effects contributing as well, although the population synthesis models suggest that these effects are likely to be small (Figure~\ref{fig:IMF_effects}b).

A least--squares fit to the SFR sub--groups of the highest redshift volume limited sample gives
\begin{equation}
\alpha \approx 0.36\log\langle SFR\rangle -2.6.
\label{math:SFR_eq}
\end{equation}

Figure~\ref{fig:IMF_slopes}(c) shows the $\alpha$ versus SFR surface density ($\Sigma_{SFR}$) relationship for the three volume limited samples. This relationship is as clean and, if anything, tighter than that seen with the specific SFR. This, again, is a consequence of accounting for galaxy size, this time through the proxy of surface area rather than mass, in refining the basic IMF--SFR relation.

A least--squares fit to $\Sigma_{SFR}$ sub--samples of the highest volume limited sample gives
\begin{equation}
\alpha \approx 0.3\log\langle \Sigma_{SFR} \rangle - 1.7.
\label{math:SFRD_eq}
\end{equation}

The models by \citet{WK05a} predict IGIMF slopes for galaxies assuming an underlying Salpeter IMF.  The resultant IGIMF slope ranges for all galaxies from these models are always steeper than the slope of the underlying IMF, with low--mass galaxies ($M_*/M_{\odot}\leq10^9$) having a steeper and wider range of IGIMF slopes ($-3.12\geq \alpha_{IGIMF}\geq-3.3$) than high--mass galaxies ($-3.07>\alpha_{IGIMF}\geq -3.1$ for $M_*/M_{\odot}\approx 10^{10}$). The general trend predicted by the IGIMF models, steep IGIMF slopes for low--mass galaxies and vice versa, seems to be in agreement with the trend found here. However, the quantitative values for the slopes predicted by the IGIMF models for a Salpeter IMF are not, and in fact a much flatter underlying IMF slope would be required to produce an IGIMF slope of around 2 \citep[see also][]{WK10}.

It is apparent that the trend to flatter IMF slopes is present as a function of each of SFR, specific SFR, and SFR surface density, although the differences between the volume--limited samples decrease most notably when SFR surface density is considered. It is reasonable, then, to infer that SFR surface density is most likely to be the underlying property on which IMF slope is primarily dependent, although the trend with specific SFR is also very tightly constrained between the different samples, and may be equally as significant. It is easier to imagine physical processes related to SFR surface density (rather than to specific SFR) that could cause variation in the IMF. For this reason we propose here that SFR surface density (or more accurately the local space density of the SFR, quantified observationally as a surface density) is the underlying property governing the solpe of the massive end of the IMF.

\section{IMF of the Milky Way and its neighbours}

The current SFR of the Milky Way is $3\pm1$ M$_{\odot}$yr$^{-1}$\,  \citep{S86}. The Spitzer/IRAC GLIMPSE survey of the Galactic plane finds a total SFR of 0.68--1.45 M$_{\odot}$yr$^{-1}$ \citep{RW10} and using the data from WMAP and GLIMPSE surveys a Galactic SFR of $1.3\pm0.2$ M$_{\odot}$yr$^{-1}$ is measured \citep{MR10}.
These SFRs are consistent with the Milky Way being placed close to the P\'EGASE model track with an input Salpeter IMF, leading to the inference of a Salpeter--like IMF for the Milky Way. Although decades of observations of the stellar IMF within the Milky Way find slopes consistent with Salpeter, this is not inconsistent with our results. The fact that external galaxies with SFRs similar to the Milky Way have similar IMF slopes is a valuable consistency check on our conclusions.  

The Milky Way does not have a constant star formation history. The Milky Way SFR was higher in the past \citep{Gilmore01}, in which case the early Milky Way would have had a flatter than Salpeter IMF (Figure~\ref{fig:IMF_slopes}). Studies of carbon--enhanced metal--poor stars report that the IMF of the early Milky Way was flatter than the present \citep{Tumlinson07, Lucatello05}, again consistent with our results. 

The measured low SFRs of the Magellanic Clouds \citep[0.14\,M$_{\odot}$yr$^{-1}$ for the Large Magellanic Cloud and 0.015\,M$_{\odot}$yr$^{-1}$ for the Small Magellanic Cloud as given by][]{MR10} and the measured IMF slopes for their stellar clusters being Salpeter or steeper \citep{Massey95} also agree with our conclusion.  In addition, the measured high SFRs ($>1000$\,M$_{\odot}$yr$^{-1}$) of submillimeter galaxies at $z>4$ \citep{MWH10} seem to require a much flatter IMF slope of $\alpha \approx -1.5$ to explain the data \citep{BLF05}, which is again consistent with our conclusion. Even more intriguingly (although perhaps only coincidentally, in particular as SFR surface density is likely to be a more fundamental relation), the extrapolation of the linear relation between $\alpha$ and $\log\langle SFR\rangle$ measured here gives $\alpha\approx-1.52$ for $SFR=1000$.

\section{Salpeter IMF vs other widely adopted IMFs}

\begin{figure}
\begin{center}
\includegraphics[scale=0.41]{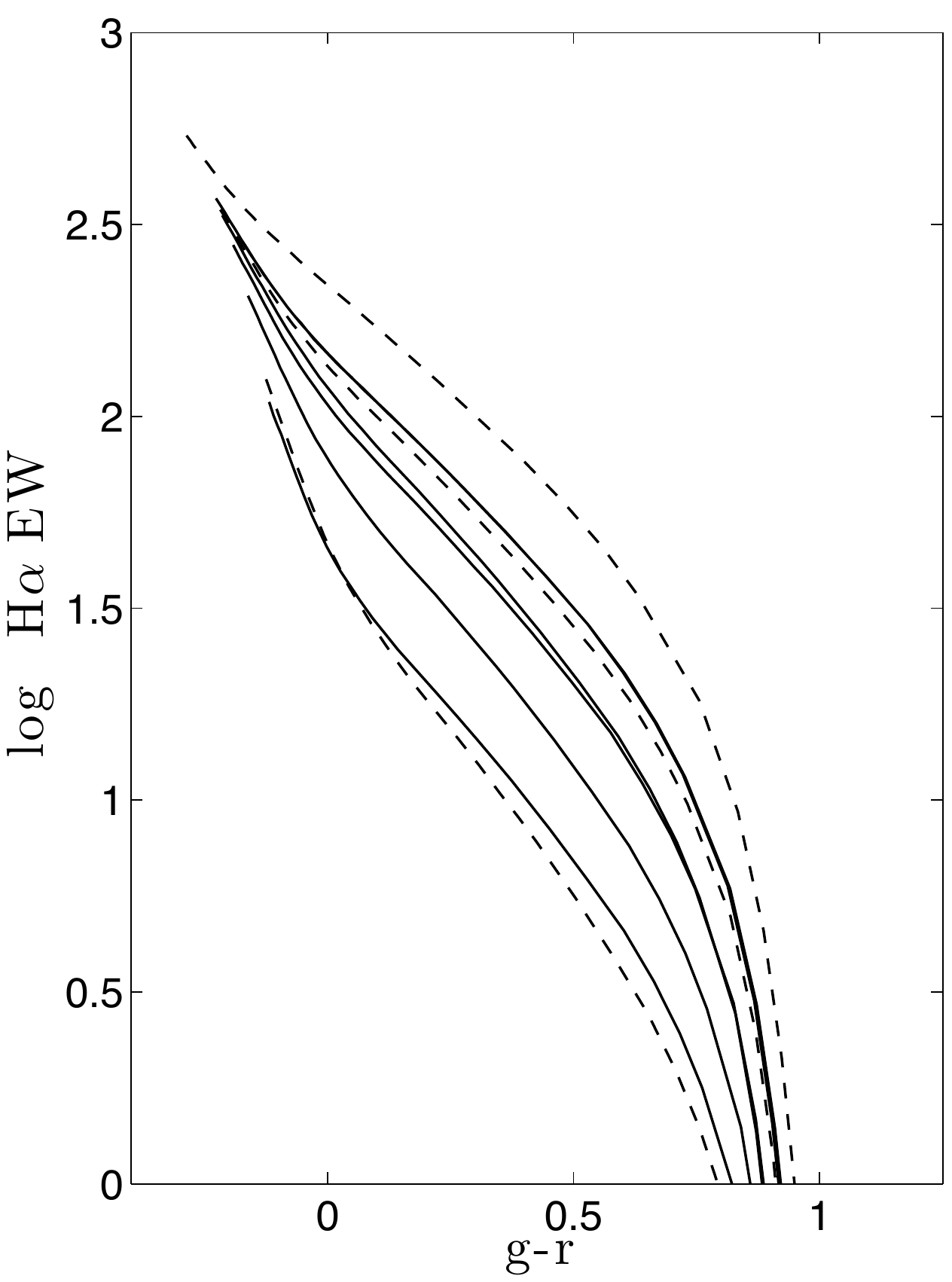}
\caption{\textit{Solid lines from top--to--bottom}: \citet{Kroupa01}A$^1$ and \citet{Kroupa01}B$^1$ IMFs produce model evolutionary tracks that overlap with one another. The rest of the evolutionary tracks are generated by using \citet{S98}, \citet{Kennicutt83}, \citet{K93} and \citet{MS79} IMFs. The dashed lines represent the three evolutionary model tracks from previous figures.}
\label{fig:commonIMFs}
\end{center}
\end{figure}

The analyses presented in this study use an IMF with a Salpeter high--mass slope. However, there are other common forms of IMF in the literature. Here we explore how the model evolutionary paths vary if an IMF with a different analytical form or slope is assumed. As Figure\,\ref{fig:commonIMFs}  shows, the commonly used IMFs in the literature produce model evolutionary paths that lie below the Salpeter IMF model track used in this study. Hence, unless the high--mass slope is adjusted appropriately, none of the common IMFs can produce an evolutionary track that describes the high SFR galaxies in the GAMA sample.  
\nocite{Kennicutt83}

\footnotetext[1]{\cite{Kroupa01}A IMF: $\alpha=-1.3$ for $0.1<M/M_{\odot}<0.5$ and  $\alpha=-2.3$ for $0.5<M/M_{\odot}<120$.\\ \indent \indent \cite{Kroupa01}B IMF: $\alpha=-1.8$ for $0.1<M/M_{\odot}<0.5$ and  $\alpha=-2.7$ for $0.5<M/M_{\odot}<1$ and $\alpha=-2.3$ for $1<M/M_{\odot}<120$.}

\section{The degeneracy with respect to turnover mass}

Here we explore the degeneracy with respect to turnover mass. The turnover mass represents the mass at which the two--part power--law IMF turns over.  

\begin{figure*}
\begin{center}
\includegraphics[scale=0.41]{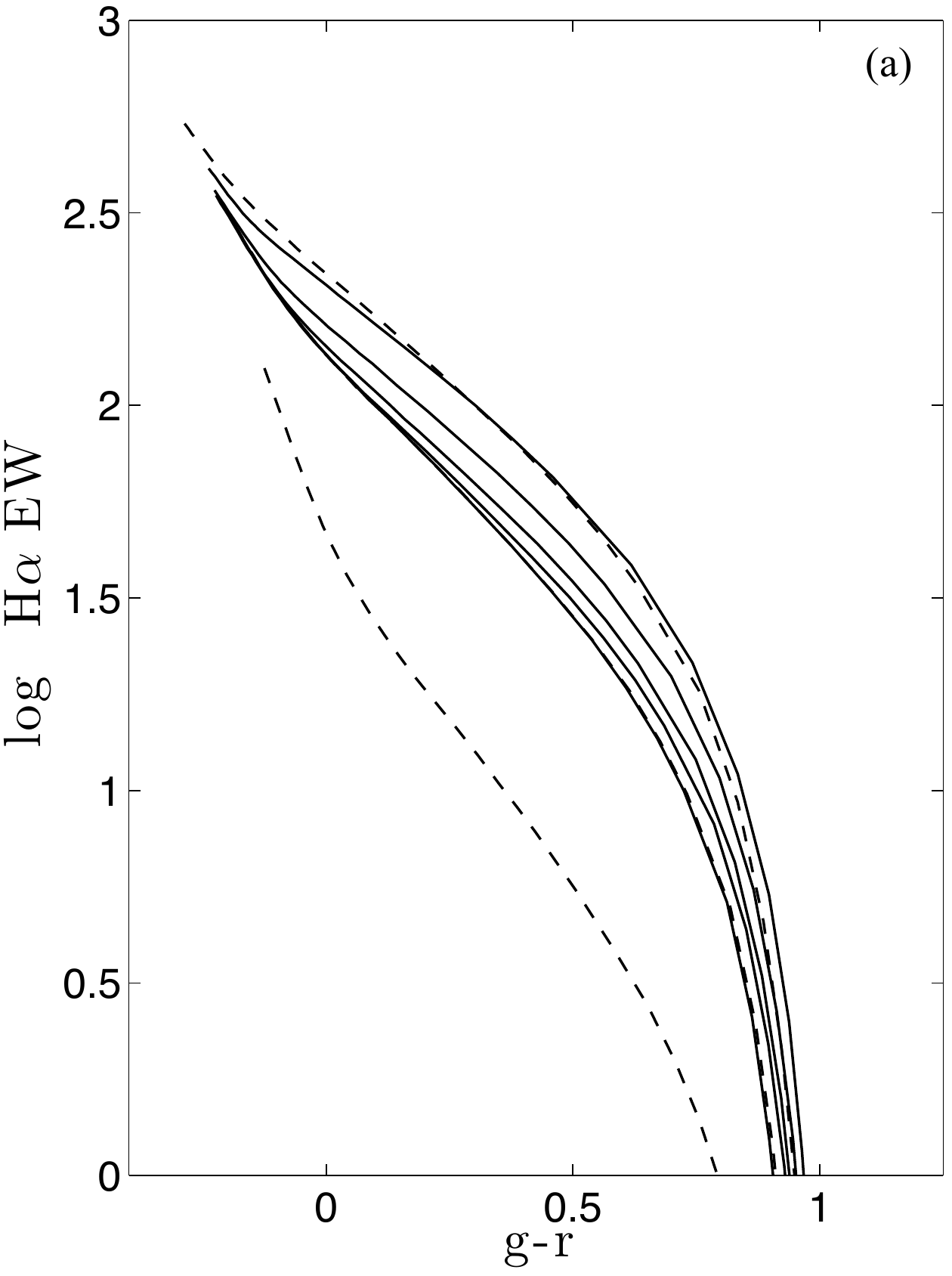}
\includegraphics[scale=0.41]{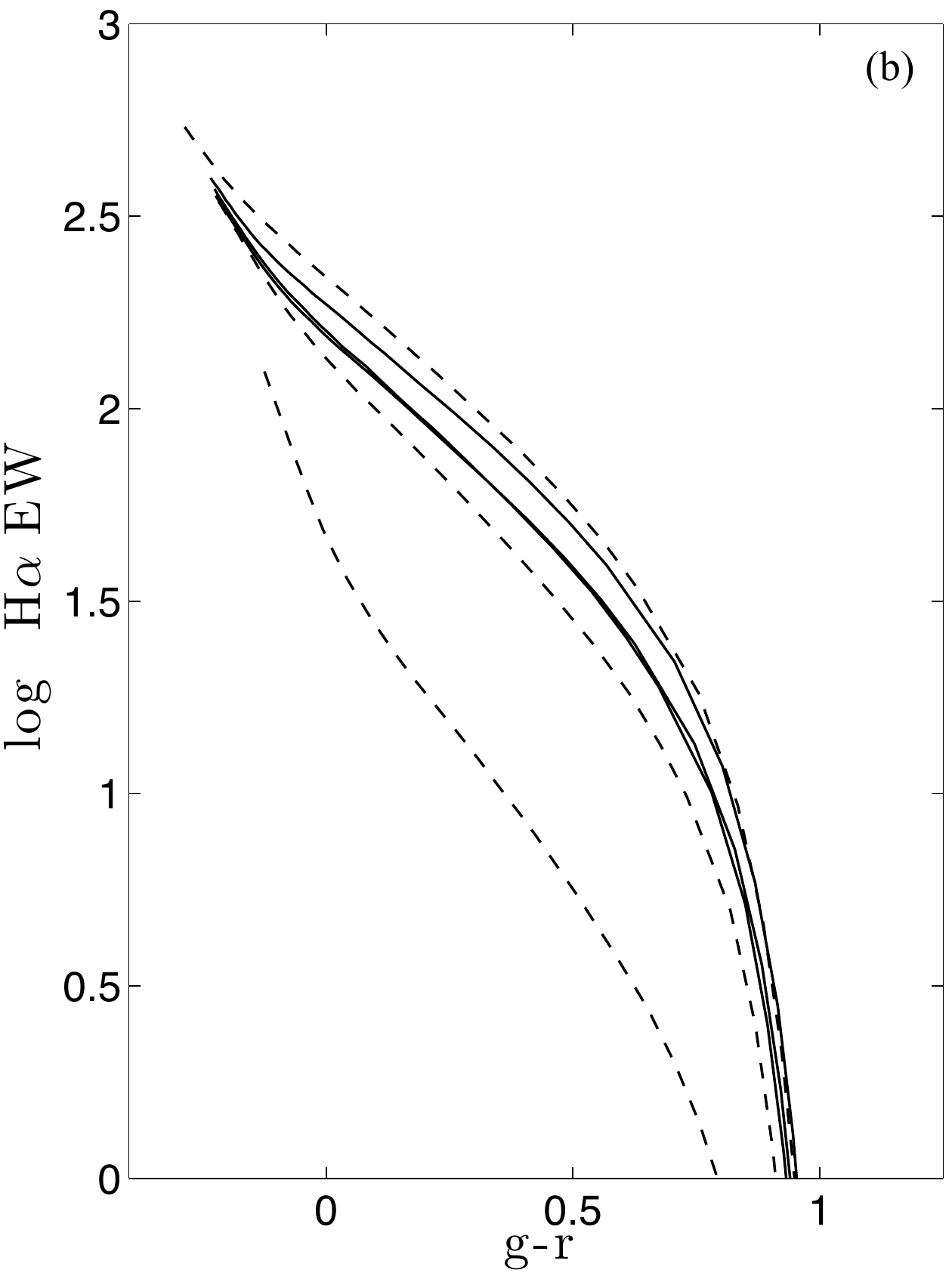}
\caption{(a) The effect of varying the turnover mass of the IMF in the case of two--part power law IMF and (b) three--part power law IMF. See the text for further details.}
\label{fig:low_mass1}
\end{center}
\end{figure*}

The solid lines (from top--to--bottom track) in Figure~\ref{fig:low_mass1}(a) are:
\begin{eqnarray}
&\alpha = &-1.3, \ \ 0.1<M/M_{\odot}<10 \nonumber \\
&& -2.35 , \ \  10<M/M_{\odot}<120. \\
&\alpha = &-1.3, \ \ 0.1<M/M_{\odot}<5  \nonumber \\ 
&& -2.35 , \ \  5<M/M_{\odot}<120. \\
&\alpha =& -1.3, \ \ 0.1<M/M_{\odot}<3  \nonumber \\
&&  -2.35 , \ \  3<M/M_{\odot}<120. \\
&\alpha =& -1.3, \ \ 0.1<M/M_{\odot}<2  \nonumber \\ 
&& -2.35 , \ \  2<M/M_{\odot}<120. \\
&\alpha =& -1.3, \ \ 0.1<M/M_{\odot}<1  \nonumber  \\ 
&& -2.35 , \ \  1<M/M_{\odot}<120.
\end{eqnarray}

The degeneracy with respect to turnover mass in the IMF is such that our results could be explained by invoking not a change in the high--mass slope of the IMF, but a progressive increase in the stellar mass at which the IMF turns over at the low--mass end. As the mass of the IMF turnover increases the model tracks shift upwards in the same manner as seen when making the high--mass slope flatter. This is illustrated in Figure~\ref{fig:low_mass1}(a), which shows that extending the low mass range up to 10 $M_{\odot}$ produces a model track that is similar to the top dashed track. 

Figure~\ref{fig:low_mass1}(b) presents the variations of the model tracks for different combinations of slopes and mass ranges of 3--part power law IMFs.  

The solid lines (from top--to--bottom track) in Figure~\ref{fig:low_mass1}(b) are:
\begin{eqnarray}
&\alpha = &-1.3, \ \ 0.1<M/M_{\odot}<0.5, \nonumber \\ 
&& -1.6 , \ \ 0.5<M/M_{\odot}<10 \textnormal{\ \ and\ \ } \nonumber  \\ 
&& -2.35 , \ \  10<M/M_{\odot}<120.\\
&\alpha = &-1.3, \ \ 0.1<M/M_{\odot}<0.5 , \nonumber \\
&& -2.0 , \ \  0.5<M/M_{\odot}<5 \textnormal{\ \ and\ \ } \nonumber \\
&& -2.35 , \ \  5<M/M_{\odot}<120.\\
&\alpha = &-1.3, \ \ 0.1<M/M_{\odot}<0.5 , \nonumber \\
 && -2.0 , \ \  0.5<M/M_{\odot}<10 \textnormal{\ \ and\ \ } \nonumber \\ 
 && -2.35 , \ \  10<M/M_{\odot}<120.
\end{eqnarray}

While it is possible that our results could be explained by a modification of the IMF such that the turnover mass increases with SFR, rather than our claimed flattening of the high--mass slope, this seems unlikely. To reproduce the high--SFR GAMA systems would require a turnover mass of $\sim10$\,M$_{\odot}$, which seems surprisingly high. This could still be a possibility, however, and is included here as an alternative explanation for completeness.   

\section{Summary}

We have used $\sim33\,000$ galaxies from the GAMA survey to confirm that the IMF does not appear to be universal. We have shown that the stellar IMF within galaxies has a strong variation with galaxy SFR.

 This result is consistent with many recent studies that suggest an evolving or varying IMF as a solution to the observed discrepancies. Many authors \citep{HB06, W08b, Wilkins08a, Fardal07, Perez08, V08, Dave08} have suggested an evolving IMF in order to reduce the discrepancy between the observed stellar mass density of the Universe and that implied by the cosmic star formation history. According to \citet{W08b, Wilkins08a} a ``cosmic'' IMF with $\alpha= -2.15$\, \citep{BG03} would solve the discrepancy between the two quantities at low redshift ($z<0.7$) but an IMF that is still flatter is required for $z>0.7$. Given that high redshift sources also tend to be the high SFR objects, our results predict the IMF of high--redshift galaxies to be very flat. Additionally, studies looking at the evolution of mass--to--light ratios and colours of galaxies \citep[e.g.][]{V08} show that the models that best fit the observations are those that assume a flatter IMF at high redshift. Galactic chemical models of \citet{Calura09} require a flatter IMF for massive galaxies in order to correctly predict their observed metallicities. We can now provide an explanation for these suggestions, with the finding that the IMF has a strong underlying dependence on the host galaxy SFR. 

\section*{Acknowledgments}
GAMA is a joint European-Australasian project based around a spectroscopic campaign using the Anglo-Australian Telescope. The GAMA input catalogue is based on data taken from the Sloan Digital Sky Survey and the UKIRT Infrared Deep Sky Survey. Complementary imaging of the GAMA regions is being obtained by a number of independent survey programs including GALEX MIS, VST KIDS, VISTA VIKING, WISE, Herschel-ATLAS, GMRT and ASKAP providing UV to radio coverage. GAMA is funded by the STFC (UK), the ARC (Australia), the AAO, and the participating institutions. The GAMA website is http://www.gama-survey.org/ . 

We thank the anonymous referee for comments that improved this paper. M.L.P.G.\ acknowledges support provided through the Macquarie University/Anglo--Australian Observatory Honours scholarship. A.M.H.\ acknowledges support provided by the Australian Research Council through a QEII Fellowship (DP0557850). We thank Prof.\,David C.\,Koo and Dr.\,Michal Michalowski for their suggestions.

\bsp

\label{lastpage}

\end{document}